# Compositional gradient engineering for enhanced ferroelectricity in ultrathin AlScN

Zekun Hu[1], Haiwen Zhang[1], Rajeev Kumar Rai[2], Yuhong Cao[1], Xiaolei Tong[1], Pedram Yousefian[1], Hyunmin Cho[1], Bongjun Choi[1], Chao-Chuan Chen[1], Yunfei He[1], Kefei Bao[1], Chloe Leblanc[1], Eric A. Stach[1,2], Roy Olsson[1], Deep Jariwala[1,*]

[1]Department of Electrical and Systems Engineering, University of Pennsylvania, Philadelphia, Pennsylvania 19104, United States of America

[2]Department of Materials Science and Engineering, University of Pennsylvania, Philadelphia, Pennsylvania 19104, United States of America

[*]Correspondence should be addressed: dmj@seas.upenn.edu (D. J.)

## Abstract

Ferroelectric AlScN is promising for CMOS-compatible non-volatile memory, but thickness scaling is limited by leakage, premature breakdown, and defect-mediated failure. Here we show that compositional grading within a continuous wurtzite AlN–AlScN lattice mitigates these limitations by distributing structural and polarization discontinuities across the film thickness, reducing defect formation and local field concentration. In a 20 nm graded heterostructure, monotonic Sc incorporation and AlN-rich boundaries produce reversible ferroelectric switching, an as-grown metal-polar state, a 21% higher breakdown field, ~10% enhanced remanent polarization, and ~40× higher resistivity relative to homogeneous AlScN. Time-domain PUND measurements reveal strongly suppressed post-switching leakage, consistent with reduced defect-assisted and polarization-coupled conduction. This improved dielectric robustness enables ferroelectric functionality in 5 nm graded stacks containing only a 2 nm $Al_{0.64}Sc_{0.36}N$ region, with measurable switching near 1 V. These results establish compositional grading as a defect- and field-management strategy for scalable ultrathin wurtzite ferroelectrics.

## Main

Ferroelectric materials enable non-volatile electronic functionality through electrically switchable polarization and underpin emerging memory and adaptive computing platforms[1, 2, 3]. Among wide-bandgap candidates, wurtzite $Al_{1-x}Sc_xN$ has attracted considerable interest owing to its large remanent polarization[4], thermal stability[5, 6], and compatibility with back-end-of-line (BEOL) integration[7]. Although lattice-matched electrodes have been explored to improve epitaxy and interface quality[8, 9], achieving low-voltage, scalable ferroelectric operation in ultrathin AlScN remains difficult.

Beyond homogeneous films, the $Al_{1-x}Sc_xN$ alloy system offers a unique structural advantage: Sc is incorporated substitutionally into the AlN wurtzite lattice, allowing the composition to be continuously varied while preserving the same crystal symmetry[10, 11]. In parallel, recent theoretical work has suggested that compositional grading in AlScN may induce proximity-driven ferroelectric switching through internal polarization coupling, highlighting the potential of spatially varying composition as a route to modulate ferroelectric stability[12]. Recent experimental work has begun to explore compositional grading in wurtzite nitrides: MBE-grown graded AlScN has been used for multi-level memory with reduced operating voltage[13], and sputtered graded films have shown large built-in fields and suppressed abnormally oriented grain growth[14] , leading to enhanced performance in radio frequency devices[15, 16]. However, none of these studies has demonstrated that compositional grading can simultaneously enhance dielectric robustness, including breakdown strength, leakage suppression, and thermal stability, while preserving ferroelectric switching, nor have they explored the ultrathin scaling limit of graded wurtzite heterostructures.

Here we show that a compositionally graded AlScN film, grown coherently from AlN on epitaxial Al to form a continuous wurtzite lattice, can simultaneously enhance both ferroelectric response and dielectric robustness, two properties that typically trade off in ferroelectric materials. Precise control of Sc incorporation during sputter deposition enables atomic-scale compositional variation across the film thickness while preserving crystallographic alignment on identical substrates. In a 20 nm heterostructure, the graded architecture yields controllable as-grown polarization and fully reversible ferroelectric switching, while simultaneously improving remanent polarization ($P_r$, +10.0%), breakdown field ($E_b$, +21.2%), leakage current, and thermal stability relative to homogeneous films. This cooperative improvement indicates that lattice-matched compositional grading can strengthen ferroelectric response while improving interfacial quality and dielectric robustness. More importantly, by smoothing structural and electrostatic discontinuities at the interfaces, this growth strategy provides a pathway to ultrathin AlScN heterostructures, achieving scaling to 5 nm stacks containing only a 2 nm $Al_{0.64}Sc_{0.36}N$ layer and supporting switching down to near 1 V.

These findings establish lattice-matched compositional grading as a viable strategy for engineering ultrathin wurtzite ferroelectrics, combining strong ferroelectric response, dielectric robustness, and scalable heterostructure design within a single crystalline framework.

## Defect-reducing growth design

Thin-film Al/ferroelectric/Al stacks were deposited on 6-inch c-plane sapphire substrates by co-sputtering Al and Sc targets (**Fig. 1a**). The Al electrodes were grown at 150 °C, while ferroelectric layers were deposited at 300 °C, maintaining a thermal budget compatible with CMOS back-end-of-line integration. Four wurtzite ferroelectric architectures with identical total thickness (20 nm) were investigated (**Fig. 1b**): a homogeneous 20 nm $Al_{0.64}Sc_{0.36}N$ film; a 15 nm $Al_{0.68}Sc_{0.32}N$ layer capped with 5 nm AlN; a 15 nm $Al_{0.68}Sc_{0.32}N$ layer grown on a 5 nm AlN seed; and a compositionally graded 17 nm $Al_{1-x}Sc_xN$ layer transitioning from pure AlN (x = 0) to $Al_{0.64}Sc_{0.36}N$, capped with a 3 nm AlN layer.

Sc incorporation was controlled by adjusting the Sc target power while maintaining a constant Al target power of 900 W, yielding the composition–thickness profiles shown in **Supplementary Fig. S1c**. In the graded structure, the 36% Sc composition change across 17 nm results in a >1% Sc variation per lattice layer, enabling atomic-scale compositional grading within a continuous wurtzite lattice.

The out-of-plane structure of the films was evaluated using θ–2θ scans around the wurtzite (0002) reflection (**Fig. 1d** and **Supplementary Fig. S1**). All samples show wurtzite nitride peaks without detectable secondary phases, confirming phase-pure growth. The AlScN-related peak position remains largely similar among the different stacks, while the additional spectral features mainly arise from the AlN-containing regions. In the AlN-seed sample, the improved crystallinity of the AlN layer allows the AlN and AlScN contributions to be more clearly resolved. By contrast, the AlN-top structure shows broader and less distinct features, suggesting reduced coherence of the top AlN layer. For the graded film, the continuously varying Sc concentration gives rise to multiple overlapping diffraction contributions rather than a single well-defined peak, consistent with a through-thickness compositional gradient.

Complementary cross-sectional scanning transmission electron microscopy (STEM) and energy dispersive spectroscopy (EDS) mapping of the graded heterostructure in **Fig. 1c** further resolve the intended layer sequence and indicate that the film is structurally continuous. The elemental distribution shows a clear spatial gradient of Sc across the film thickness, while the Al-containing layers remain laterally continuous, consistent with the designed graded AlScN architecture. The complete set of elemental maps is presented in **Supplementary Fig. S2**, and these further corroborate the preserved stack integrity and the intended through-thickness compositional gradient.

For the ultrathin stack, the cross sectional ADF STEM image in **Fig. 1e** shows a continuous 5 nm active nitride layer embedded between the Al electrodes, following the designed sequence of a 2 nm graded AlN to AlScN region, a 2 nm $Al_{0.64}Sc_{0.36}N$ region, and a 1 nm AlN top cap. The atomic resolution HAADF STEM image in **Fig. 1f** further reveals ordered lattice contrast across the ultrathin nitride layer, indicating that crystallinity is preserved even after aggressive thickness scaling. Together with the uniform as-grown M-polar state observed in the ultrathin graded stack, these results indicate well controlled pristine polarity, stable nucleation, and preserved structural integrity in the ultrathin regime. The complete elemental maps in **Supplementary Fig. S3** further corroborate the preserved stack integrity, lateral continuity of the Al containing layers, and intended compositional gradient across the film thickness.

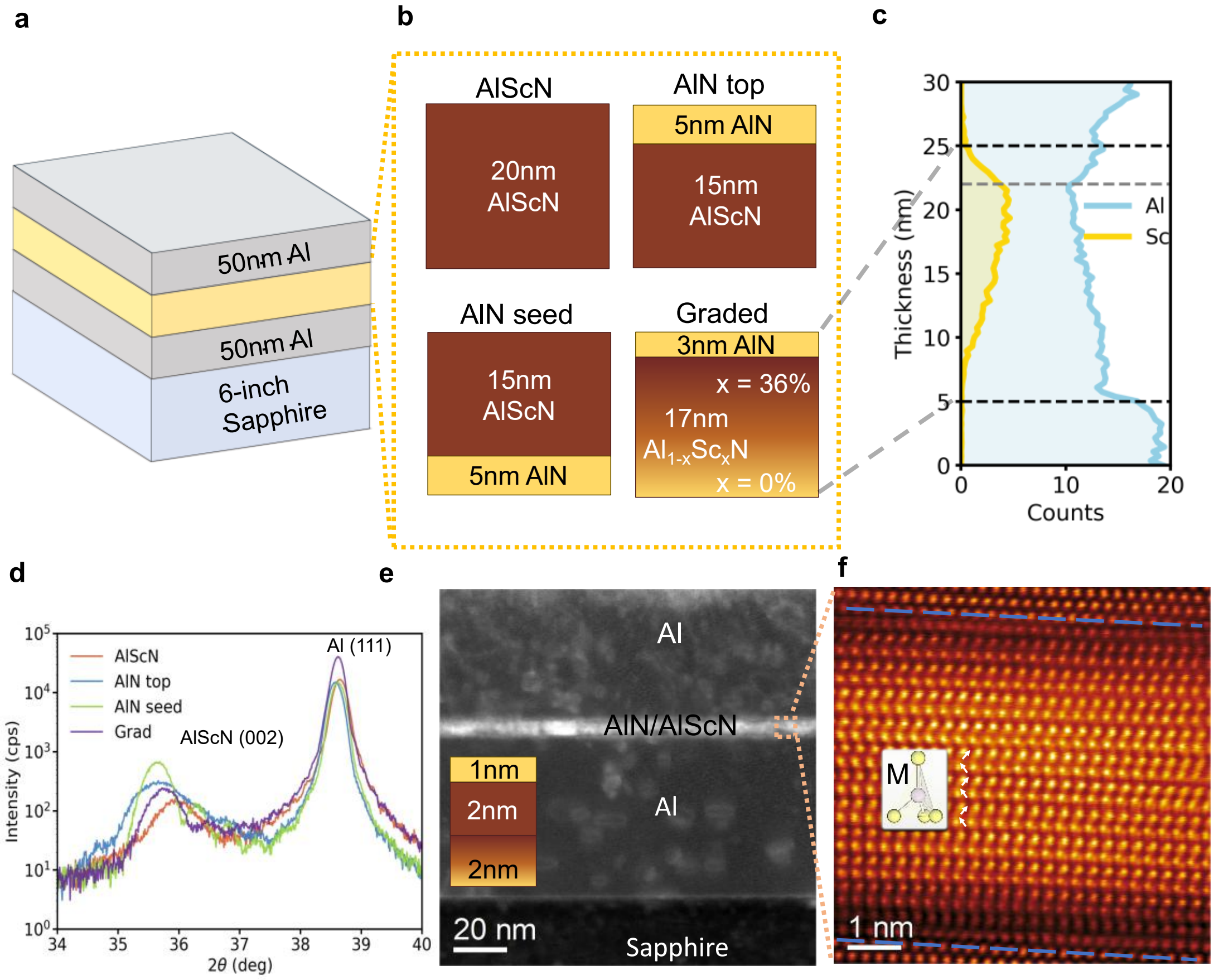

**Fig. 1. Design and structural characterization of the graded AlScN heterostructures. a,** Schematic of the fully in situ grown Al/ferroelectric/Al stack on c-plane sapphire, consisting of a 20 nm active nitride layer sandwiched between 50 nm Al electrodes. **b,** Heterostructure designs investigated in this work: homogeneous AlScN, AlScN with an AlN top constraint, AlScN on an AlN seed layer, and graded $Al_{1-x}Sc_xN$ with $0 \le x \le 0.36$. **c,** STEM EDS line profiles of Al and Sc across the graded heterostructure, confirming the intended through thickness compositional grading and the AlN rich top region. **d,** θ–2θ X-ray diffraction scans showing the wurtzite AlScN (0002) and epitaxial Al (111) reflections, consistent with single-phase growth and the absence of detectable secondary phases. **e,** Cross sectional HAADF STEM image of the 5 nm graded heterostructure integrated between Al electrodes on sapphire. The ultrathin stack consists of a 2 nm graded AlN to $Al_{0.64}Sc_{0.36}N$ region, a 2 nm $Al_{0.64}Sc_{0.36}N$ layer, and a 1 nm AlN top cap. **f,** Atomic resolution HAADF STEM image from the region highlighted in **e**, showing a structurally continuous wurtzite lattice across the ultrathin graded heterostructure, with the atomic arrangement consistent with M-polar ordering.

## Controllable intrinsic polarization

Previous efforts to control metal-polar (M-polar) and nitrogen-polar (N-polar) orientations in wurtzite nitride films have largely focused on substrate and buffer-layer engineering[17, 18]. In sputtered AlScN, the as-grown state is typically reported to be N-polar, although composition[19] and ion energy[20] may also affect the preferred polarization orientation (**Supplementary Table S1**). In contrast, our results show that film polarity can be deterministically controlled through heterostructure design, even on the same substrate.

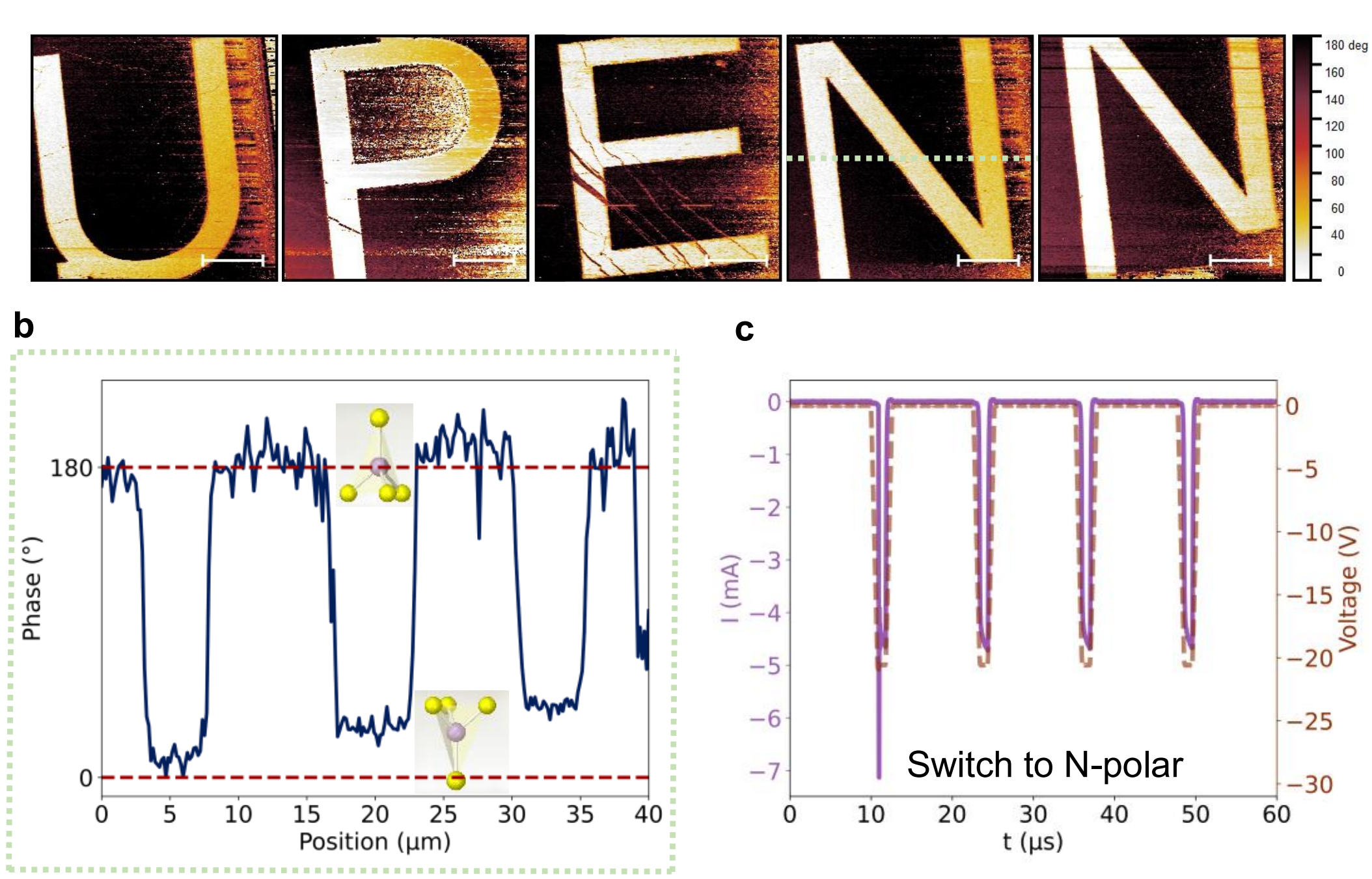

**Fig. 2. Intrinsic polarization orientation and deterministic switching in graded AlScN. a,** PFM phase images of the graded film. Regions patterned into the "UPENN" geometry are locally poled to N-polar, while the surrounding area retains the as-grown polarization state. Scale bar, 10 μm. **b,** PFM phase line profile across the poled boundary (dashed line in a), showing a ~180° phase contrast between domains, consistent with full polarization reversal. **c,** Time-resolved switching under sequential negative voltage pulses. The first pulse produces a displacement-current transient, indicating reversal from the initial as-grown polarization orientation.

To probe the initial polarization state of the graded film, Al top electrodes were patterned into the letters "UPENN", driven with negative pulses exceeding the switching voltage, removed by wet etching, and subsequently examined by PFM. While the pure AlScN reference film shows an as-grown N-polar state (**Supplementary Fig. S4**), the graded film exhibits an as-grown M-polar state, as evidenced by the PFM phase images and line profile in **Fig. 2a,b** and further supported by the atomic resolution HAADF STEM image in **Fig. 1f**. The approximately 180° phase contrast between the switched and pristine regions indicates a homogeneous initial M-polar orientation across the 20 nm graded film. Correspondingly, the first negative pulse in **Fig. 2c** produces a clear switching current, while later pulses do not, consistent with polarization reversal occurring during the initial pulse.

The initial M-polar state is unlikely to arise from the compositional gradient alone. Recent STEM measurements[14] on thick sputtered graded AlScN showed N-polarity even in films with large gradient-induced built-in fields, indicating that internal fields from grading do not necessarily determine the pristine polar orientation. In our samples, both AlN-seed and AlN-top control stacks also show M-polarity, whereas the uniform AlScN reference is N-polar. This comparison suggests that AlN-rich interfacial boundary conditions are the dominant factor in polarity selection, while the through-thickness gradient provides an additional internal chemical and polarization asymmetry. Thus, the graded film should be viewed as an interface-coupled graded heterostructure rather than a purely gradient-driven polarity switch.

## Enhanced ferroelectricity through compositional grading

We next evaluate how compositional grading affects ferroelectric switching and electrical robustness using the via-defined capacitor geometry shown in **Fig. 3a**. This structure preserves the as-grown top Al electrode and avoids direct mechanical loading during probing, enabling a more intrinsic comparison among the four heterostructures. Details of the device design and fabrication are provided in the Methods and **Supplementary Fig. S5**.

The graded heterostructure shows a distinct combination of suppressed leakage and robust ferroelectric switching. In the DC J–V characteristics (**Fig. 3b**), the homogeneous AlScN film exhibits pronounced polarization-dependent conduction, whereas the AlN-top and AlN-seed structures show asymmetric leakage suppression due to the insertion of an AlN barrier at one interface[10]. By contrast, the graded stack, which incorporates AlN-rich boundaries on both sides together with a continuous compositional transition, exhibits the lowest leakage current among the four structures. This reduced conduction suggests that the graded design suppresses charge injection and defect-assisted transport[21] while maintaining ferroelectric functionality. Additional transport analysis is provided in **Supplementary Figs. S6 and S7**.

Dynamic AC J–V measurements at 100 kHz further confirm reversible polarization switching in all four stacks, as evidenced by clear switching-current peaks (**Fig. 3c**). The frequency dependence of coercive field and device-to-device variation are presented in **Supplementary Figs. S8** and **S9**. Complementary C–V measurements, plotted as the capacitance change relative to the minimum-capacitance point to account for via-related parasitics, show clear bias-induced capacitance transitions below 10 V for all samples under quasi static DC bias, while capacitance was measured using a small 100 kHz AC signal (**Fig. 3d**). The graded film displays a more restrained and stable capacitance evolution, consistent with improved dielectric robustness during field cycling.

PUND (Positive-Up–Negative-Down) measurements were conducted to evaluate the remanent polarization of the four stacks (**Fig. 3e**). With a voltage resolution of 0.1 V, the polarization difference was clearly resolved. The graded film shows well-defined saturation on both switching branches, whereas the AlN-seed and AlN-top structures saturate on only one branch each, and the homogeneous AlScN film does not exhibit similarly clear saturation within the measured window. This more symmetric behavior is consistent with the bilateral AlN-rich interfaces and continuous composition profile in the graded heterostructure, which together help balance carrier injection and switching response. The corresponding tunnelling interface schematics are shown in **Supplementary Fig. S10**, and the full PUND responses and device variations of 3 devices are provided in **Supplementary Fig. S11**. Extracting the saturated $2P_r$ values shows that the graded film reaches approximately 123 μC cm$^{-2}$, about 10% higher than the other three stacks (110–111 μC cm$^{-2}$). We attribute this enhancement to the relatively lower Sc concentration in the effective switching region of the graded AlScN[22].

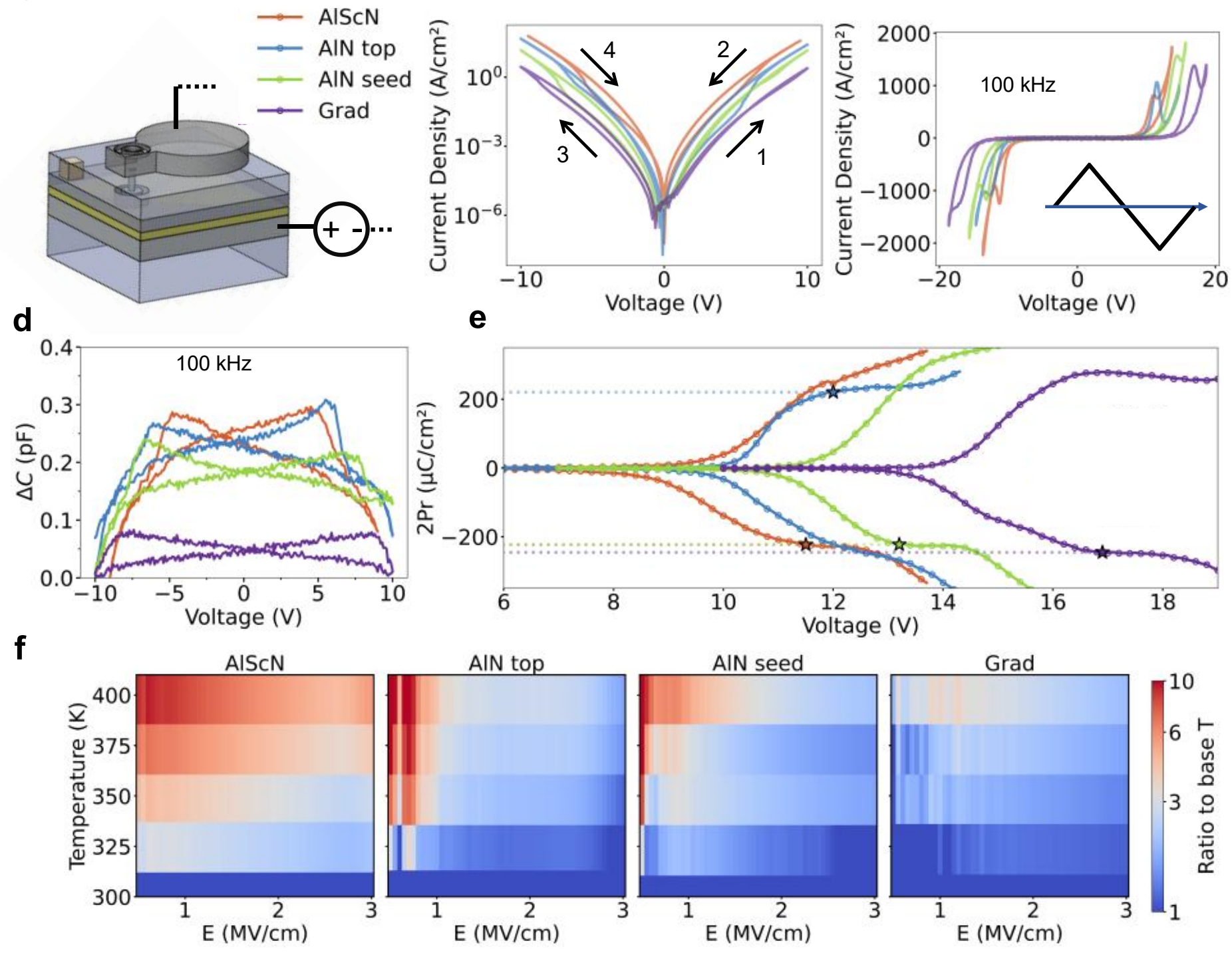

**Fig. 3. Ferroelectric switching across engineered AlScN stacks. a,** Via-defined capacitor geometry with a 5 μm radius enabling electrical access while preserving the as-grown top Al electrode and minimizing extrinsic strain. **b,** DC J–V characteristics showing polarization-dependent conduction for all architectures. **c,** High-frequency (100 kHz) dynamic switching loops, confirming robust polarization reversal under fast excitation. **d,** ΔC–V response at 100 kHz, exhibiting characteristic butterfly behavior associated with ferroelectric domain switching. **e,** PUND measurements with 0.1 V step resolution. Extracted saturated $2P_r$ values (stars) quantify remanent polarization independent of leakage contributions. **f,** Temperature- and field-dependent switching maps (ratio to baseline), illustrating enhanced temperature stability in the graded structure.

The switching stability was further examined from 25 °C to 125 °C, where the graded stack exhibits the smallest variation with temperature among the four structures (**Fig. 3f** and **Supplementary Fig. S12**). Combined with its substantially lower leakage current, this behavior points to the potential of the graded heterostructure for harsh-environment electronics[23]. Further development of electrode materials and device integration will be needed to translate this intrinsic materials advantage into a practical platform for extreme-environment operation, as the present devices employ Al electrodes.

## Compositional gradient suppresses polarization-activated leakage pathways

**Fig. 4** provides direct time-domain evidence that the PUND response of these ultrathin AlScN-based stacks contains both intrinsic ferroelectric switching and a polarization-coupled leakage component. While all four films show clear switching transients during the P and N pulses, the steady-state current extracted from the boxed regions after reversal varies markedly across the different architectures. Consistent with previous reports on ultrathin AlScN[9], the homogeneous AlScN capacitor exhibits the largest post-switching leakage (**Fig. 4b**), whereas insertion of AlN (**Fig. 4c** and **4d**) at either interface suppresses this residual conduction, and the graded structure yields the lowest steady-state current overall (**Fig. 4e**). These results indicate that the apparent hysteretic response is shaped not only by polarization reversal, but also by dynamic transport processes activated during switching[24, 25]. Within this picture, the AlN-modified interfaces reduce charge injection and suppress the electrical activation of these transient channels, while the graded stack achieves the strongest suppression by engineering both interfaces simultaneously.

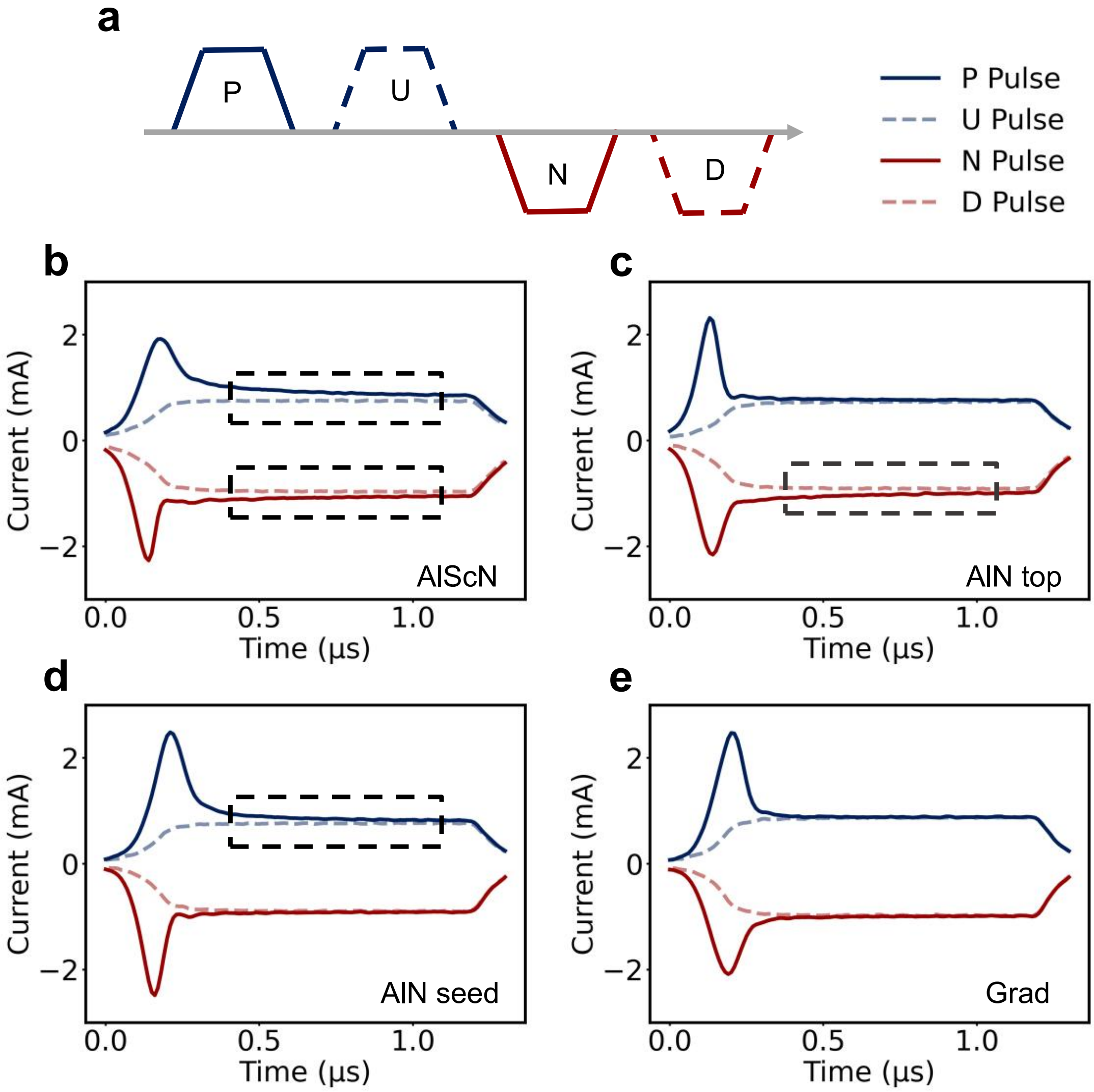

**Fig. 4. Interface-controlled post-switching leakage in AlScN films.**
**a,** PUND pulse sequence (P, U, N, D) used to resolve switching transients and subsequent steady-state conduction. **b–e,** Time-resolved current responses for AlScN (b), AlN-top (c), AlN-seed (d), and graded (e) films. Solid traces correspond to switching pulses (P, N); dashed traces to reference pulses (U, D).
The boxed regions highlight the steady-state regime following polarization reversal. Distinct leakage levels appear after switching, indicating polarization-dependent conduction that is strongly modulated by the AlN interfaces. The graded film exhibits the lowest post-switching leakage, consistent with interface-controlled transport.

The complete 100 kHz triangular PUND datasets and the corresponding extracted P–E loops are provided in **Supplementary Figs. S13 and S14**. In the extracted loops, the additional non-switching charge contribution, highlighted by the grey shaded regions, is clearly suppressed on both switching branches in the graded structure, further confirming that interface engineering reduces leakage-related current contributions during polarization reversal.

## Ultrathin films enabled by reduced defect sensitivity

A key question is whether the same design principle can survive in the ultrathin limit, where AlScN typically faces stronger leakage, a narrower switching window, and a rapidly increasing coercive field. Prior work has shown that AlScN can remain ferroelectric near 5 nm, including sputtered films at 5.4 nm[26], sub-5 nm films[9, 27] with reported switching at 1 V, and ultrathin films enabled by coherent HfN electrodes[8]. Together, these studies show that ultrathin AlScN ferroelectricity is achievable, but also highlight the central role of interfacial and structural design in stabilizing switching at the few-nanometer scale.

We therefore applied the graded strategy to a family of ultrathin stacks (**Supplementary Fig. S15**): an 8 nm graded film with a 5 nm $Al_{0.64}Sc_{0.36}N$ region, a 6 nm symmetric AlN/AlScN/AlN stack with a 4 nm $Al_{0.64}Sc_{0.36}N$ layer, a 6 nm graded film with a 4 nm $Al_{0.64}Sc_{0.36}N$ region, and a 5 nm graded film containing only a 2 nm $Al_{0.64}Sc_{0.36}N$ layer. In the graded structures, the bottom 2 nm was compositionally graded from AlN to $Al_{0.64}Sc_{0.36}N$, while the top interface remained capped by 1 nm AlN. This architecture preserves AlN-rich boundaries while reducing the thickness of the high-Sc region to the few-nanometer limit.

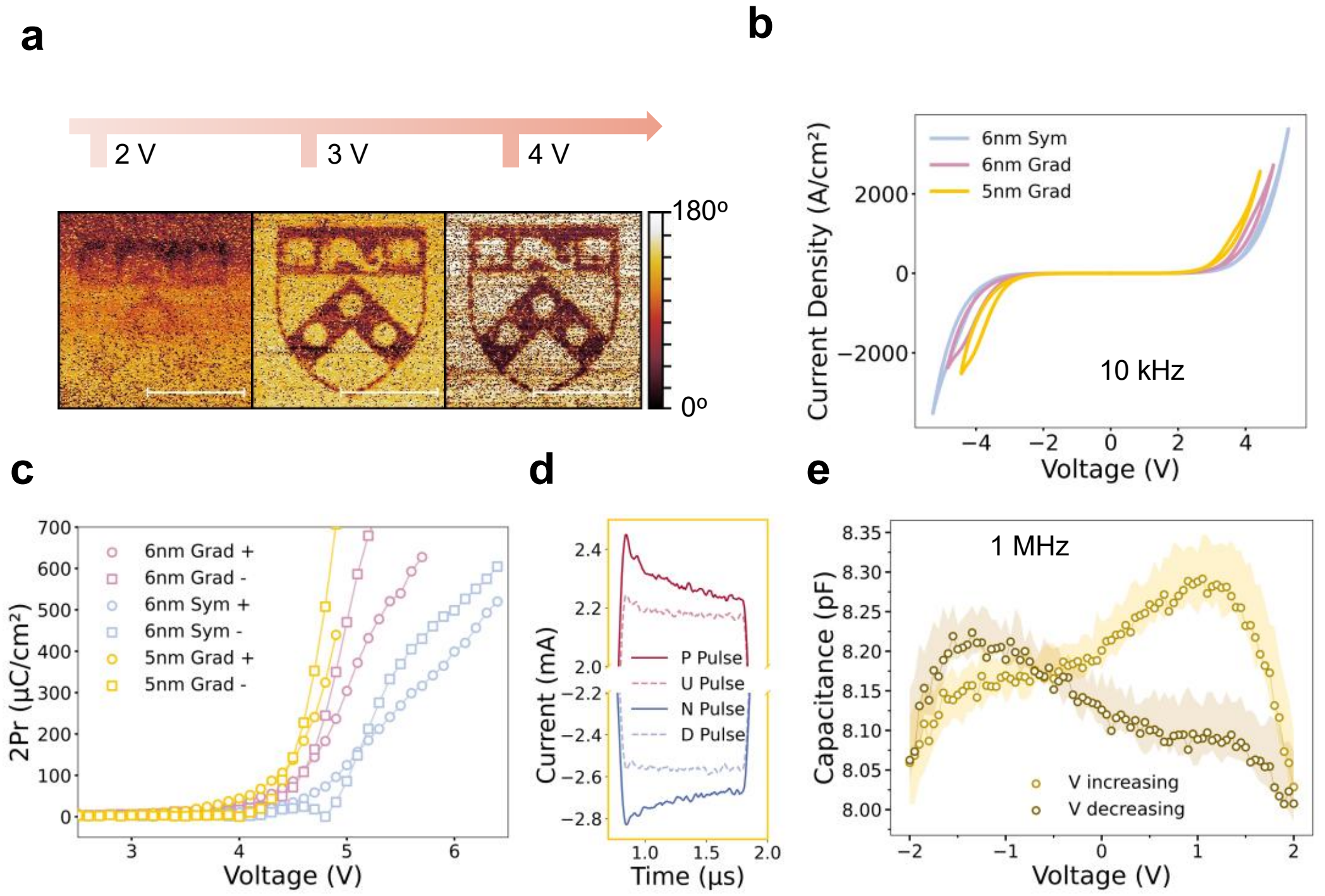

**Fig. 5. Ultrathin ferroelectric functionality enabled by compositional grading.**
**a,** PFM tip-bias switching of the 8 nm graded AlScN heterostructure. Increasing tip bias from 2 to 4 V induces gradual polarization reversal. Scale bar, 10 µm. **b,** AC current density–voltage characteristics measured at 10 kHz for the 6 nm symmetric AlN-stack reference, 6 nm graded film, and 5 nm graded film. **c,** Gradually increasing PUND measurements for the 6 nm graded, 6 nm symmetric, and 5 nm graded films. **d,** Time-resolved PUND current response of the 5 nm graded film, showing distinct switching and non-switching pulses (P, U, N, and D). **e,** C–V characteristics of the 5 nm graded film with a 10 µm radius, with shaded regions indicating device-to-device variation across n = 10 capacitors.

PFM tip-bias writing on the 8 nm graded film reveals progressive switching as the bias is increased from 2 to 4 V (**Fig. 5a**). Rather than an abrupt binary transition, the switched contrast develops gradually, indicating multistate polarization control in the ultrathin graded heterostructure.

Electrical measurements confirm that switching signatures persist at still smaller thickness. The 10 kHz AC J–V curves show clear switching peaks for both the 6 nm and 5 nm graded stacks, in comparison with the 6 nm symmetric reference (**Fig. 5b**). Notably, the 5 nm graded AlScN heterostructure remains switchable under an 80 ns voltage pulse, with an estimated write energy of ~83 pJ/um² per pulse (**Supplementary Fig. S16** and **Note S2**). In the gradually increasing PUND measurement, the extracted $2P_r$ rises systematically with pulse amplitude for the graded films, including the 5 nm stack (**Fig. 5c** and **Supplementary Fig. S17**), indicating field-driven and reversible switching rather than purely leakage-dominated transport. The 5 nm graded device also exhibits distinct P, U, N, and D responses in the time domain (**Fig. 5d**), although the extracted $2P_r$ still includes some post-switching leakage contribution.

The C–V characteristics of the 5 nm graded film exhibit a reproducible hysteretic response across 10 devices (**Fig. 5e**), demonstrating switching near 1 V. Capacitance was measured using a 1 MHz small AC signal while the DC bias was swept quasi-statically. Taken together, these results show that compositional grading does not merely improve 20 nm films, but provides a viable route for extending ferroelectric functionality into the ultrathin regime. Even when the $Al_{0.64}Sc_{0.36}N$ region is reduced to only 2 nm, the graded heterostructure retains measurable switching and hysteretic dielectric response. We attribute this robustness to AlN-rich interfaces that suppress carrier injection and compositional grading that distributes lattice distortion and polarization discontinuities across the film thickness, thereby reducing defect formation, local field concentration, and defect-assisted leakage in ultrathin films. In this sense, graded engineering provides a pathway to high-quality ultrathin AlScN films, complementing prior approaches based on electrode selection and multilayer interface engineering.

## Reliability and dielectric performance

As shown in **Fig. 6a**, the 20 nm graded heterostructure delivers a clear improvement in the key performance metrics relative to the AlScN control. The maximum breakdown field, extracted from 30 devices (n = 30), increases by 21.2% to 12.3 MV/cm, reflecting a substantial gain in dielectric reliability, while the DC coercive field (n = 10) rises by 52.2%. Together with the approximately 10% enhancement in remanent polarization, these changes point to a larger and more robust memory window for future transistor integration[28]. The endurance, presented in **Supplementary Fig. S18**, is presently comparable across the structures, although further optimization should enable additional gains in cycling stability[29]. The graded stack also shows a 17% lower permittivity, reaching 13.3 as extracted from area-dependent capacitance measurements, and an approximately 40-fold increase in resistivity at 4 V DC (n = 10).

The breakdown-voltage distributions in **Fig. 6b** (Weibull plots in **Supplementary Fig. S18**) were measured on devices using 100 kHz AC poling, with the voltage increased in 0.1 V steps until breakdown. A total of 30 devices were tested for each stack to ensure robust statistics. Notably, the graded film shows a more concentrated distribution than the other structures, indicating reduced variability across devices.

**Fig. 6c** benchmarks the engineered AlScN heterostructures against a wider set of dielectric and ferroelectric materials by plotting breakdown field against permittivity[30, 31, 32, 33, 34, 35, 36, 37, 38, 39, 40]. The graded structure occupies a favorable region of this design space, combining relatively low permittivity and high breakdown strength without sacrificing ferroelectricity. This is notable because high-breakdown dielectrics are typically non-switchable, whereas ferroelectric materials often trade dielectric robustness for polarization functionality. In contrast, the 20 nm graded AlScN film retains strong remanent polarization while approaching the electrical resilience expected from advanced dielectric layers. Ultraviolet ellipsometry further confirms the optical bandgaps of the AlScN and graded films, as presented in **Supplementary Fig. S19–S21** and **Note S1**. Beyond this materials-level advantage, AlScN also offers a distinct technological benefit through CMOS-compatible processing and intrinsic scalability in ultrathin nitride heterostructures. Together, these results suggest that engineered wurtzite ferroelectrics may enable ferroelectric platforms in which polarization switching is retained alongside the high-field tolerance and low-loss characteristics needed for next-generation electronic integration.

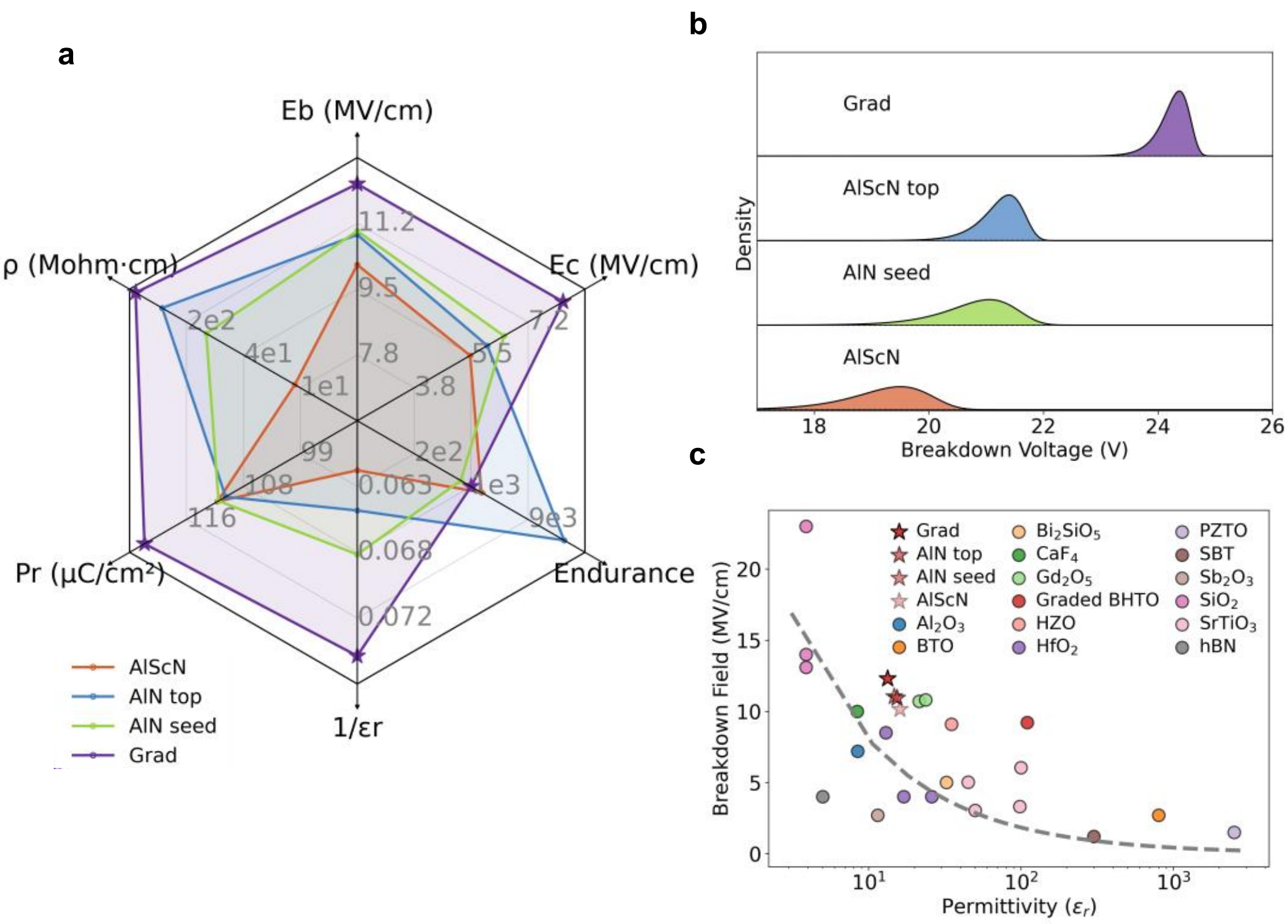

**Fig. 6. Reliability and dielectric scaling advantage of graded AlScN.**
**a,** Multi-parameter comparison of 20 nm AlScN, AlN-top, AlN-seed, and graded films, including breakdown field (Eb), coercive field (Ec), endurance, inverse permittivity ($1/\varepsilon_r$), remanent polarization (Pr), and resistivity (ρ). **b,** Breakdown voltage distributions measured across n = 30 devices per film. **c,** Benchmark comparison of breakdown field versus permittivity across representative dielectric and ferroelectric materials.

## Discussion and Conclusion

The central finding of this work is that compositional grading within a single wurtzite lattice can resolve the long-standing trade-off between ferroelectric switching and dielectric robustness. In conventional homogeneous AlScN, the abrupt polarization discontinuity at the metal-ferroelectric interface creates a large depolarization field that promotes charge injection and defect-assisted leakage. By replacing this abrupt interface with a continuous AlN→AlScN compositional transition, the graded heterostructure distributes the polarization discontinuity across a finite thickness, reducing the local electric field at each lattice plane. This mechanism is analogous to the field-spreading effect observed in compositionally graded perovskite ferroelectrics[41, 42], where continuous variation in polarization magnitude suppresses domain-wall pinning and reduces coercive-field inhomogeneity. In the wurtzite system, the same principle operates through the monotonic increase in spontaneous polarization with Sc content: AlN-rich boundary layers present a smaller polarization mismatch with the metal electrode, while the gradient region accommodates the full polarization change over many unit cells rather than at a single interface. This distributed transition may reduce the formation or electrical activation of misfit related point defects, including nitrogen vacancy related charged defects, which can contribute to Poole Frenkel leakage in nitride ferroelectrics[43, 44]. The improved breakdown strength, suppressed post-switching leakage, and enhanced resistivity observed in the graded films are all consistent with a lower effective defect density in the interfacial region. The observation that grading also inverts the as-grown polarity from N-polar to M-polar further supports the role of the AlN seed region in templating the initial growth polarity, consistent with the preferential M-polar nucleation reported for AlN on metallic substrates[19]. This more gradual structural and electrostatic transition can reduce defect formation, local field concentration, and defect-assisted leakage pathways. Accordingly, the graded heterostructure exhibits lower leakage, improved remanent polarization, enhanced thermal stability and higher breakdown strength than the homogeneous AlScN reference.

A notable implication is that the graded architecture enables ultrathin ferroelectric functionality without relying on exotic electrode materials or epitaxial substrates. The demonstration of switching in a 5 nm stack containing only 2 nm of $Al_{0.64}Sc_{0.36}N$ achieved with standard sputtered Al electrodes on sapphire suggests that the graded design itself is the primary enabler of ultrathin operation. Looking forward, the endurance of graded stacks, which is presently comparable to that of homogeneous films, may benefit from the reduced defect density if the dominant fatigue mechanism proves to be defect accumulation rather than field-driven nucleation. Phase-field modelling of switching dynamics in graded wurtzite lattices, combined with depth-resolved defect spectroscopy, would clarify these mechanisms. More broadly, these findings highlight lattice-matched compositional grading as a general heterostructure strategy for stabilizing ferroelectricity, suppressing leakage, and extending wurtzite ferroelectrics toward their ultimate thickness-scaling limit.

# Methods

## Growth of ferroelectric films

The AlScN thin film with a scandium concentration of 36% was grown on a (111)-oriented 50 nm Al layer deposited on c-axis oriented sapphire wafer using a physical vapor deposition (PVD) system modified from previous work [7, 41]. First, a 50 nm Al layer was deposited by DC sputtering at 150 °C in an Evatec CLUSTERLINEVR 200 II pulsed DC PVD system with an Ar gas flow of 20 sccm. Without breaking vacuum, the AlScN thin film was then deposited by co-sputtering from separate Al and Sc targets, with a power of 900 W for Al and 555 W for Sc. Both targets had a diameter of 100 mm. The deposition was carried out at 300 °C under a vacuum of $3 \times 10^{-4}$ Torr, with an $N_2$ flow of 30 sccm. The graded film was deposited by linearly increasing the Sc target power to 555 W. The AlN film was deposited at 300 °C under an $N_2$ flow of 21 sccm. Following the AlScN/AlN deposition, a 50 nm Al capping layer was deposited in situ without breaking the vacuum, effectively preventing surface oxidation of the AlScN film and preserving its ferroelectric properties.

## Device Fabrication

Ti/Au alignment markers were first deposited on the Al/AlScN/Al wafers using a Lesker PVD75 e-beam evaporator. Device isolation was then defined by electron beam lithography with a Raith EBPG5200+ system and dry etching (Oxford Cobra ICP) of the in-situ top Al layer. A 200 nm $SiO_2$ layer was deposited by PECVD and patterned by reactive ion etching using an Oxford 80 Plus RIE system to form device vias. After resist removal, the native oxide on the exposed Al was removed by RF Ar sputter etching in an Explorer 14 sputterer at 200 W for 10 min, followed immediately, without vacuum break, by sputter deposition of 230 nm Al to fill the vias. A final lithography and ICP dry etch step using an Oxford Cobra ICP system was used to define and isolate the contact pads.

## Electrical characterization

Electrical measurements, including DC and AC current–voltage characterization, PUND, and endurance tests, were carried out using a Keithley 4200A-SCS parameter analyzer with the NVM library. For all measurements, bias was applied to the bottom electrode while the top electrode was grounded. Unless otherwise stated, PUND measurements were performed with a pulse width of 1 μs, a rise/fall time of 0.8 μs, and an inter-pulse delay of 10 μs. In both DC and AC measurements, the voltage sweep was first applied in the positive direction and then in the negative direction. Prior to data acquisition on the thin-film devices, wake-up AC poling was performed until stable switching behavior was observed. Unless otherwise noted, all devices used for electrical characterization were circular capacitors with a radius of 5 μm.

## TEM Characterization

Cross-sectional STEM and EDS were performed using a probe-corrected JEOL NEOARM equipped with two large-angle solid state SDD detectors for EDS measurement. The high-angle annular dark-field (HAADF)-STEM images and EDS maps were acquired at operating voltage of 200 kV with 25 mrad convergence angle, a camera length of 6 cm and dwell time of 16 ms per pixel. The sample was prepared using a FEI Strata DB235 dual beam focused ion beam / scanning electron microscope.

## PFM Characterization

PFM measurements were performed on a commercial Asylum Research system using Pt/Ir-coated silicon probes operated at a contact-resonance frequency of 75 kHz. Tip-bias PFM images were acquired in Litho-PFM mode by first writing with the tip bias and then reading the resulting domain pattern.

## XRD Characterization

Structural characterization was carried out using a Rigaku SmartLab SE X-ray diffractometer equipped with a Cu Kα source. A Ge(220) monochromator was employed for all scans. The θ–2θ patterns of all samples exhibited distinct AlScN (002) reflections together with Al (111) peaks, suggesting the epitaxial growth of (002)- textured wurtzite AlScN films on Al (111) bottom electrodes.

## Acknowledgements

We thank Mark Qu at Johns Hopkins University for assistance with data collection. We also acknowledge the staff of the Quattrone Nanofabrication Facility at the University of Pennsylvania for their support and helpful suggestions.

### Funding:

D.J. and Z.H. acknowledge primary support from the Office of Naval Research (ONR) Nanoscale Computing and Devices program (No. N00014–24–1–2131). D.J. and Z.H. also acknowledge support from the Air Force Office of Scientific Research (AFOSR) GHz-THz Program (Grant No. FA9550–23–1–0391). D.J. and Z. H. acknowledge support from the National Science Foundation, Future of Semiconductors (FuSe) program: award number 2328743. A portion of the sample fabrication, assembly, and characterization were carried out at the Singh Center for Nanotechnology at the University of Pennsylvania, which is supported by the National Science Foundation (NSF) National Nanotechnology Coordinated Infrastructure Program grant NNCI-1542153. The authors acknowledge the use of an XRD facility supported by the Laboratory for Research on the Structure of Matter and the NSF through the University of Pennsylvania Materials Research Science and Engineering Center (MRSEC) DMR-2309043.

#### Author Contributions:

D.J., R.O., Z.H. and H.C. conceived and designed this work. Z.H. and H.Z. carried out all the electrical measurements. Z.H., C.-C.C. and Y.H. fabricated the devices. Z.H. and H.Z. analyzed the electrical measurements. D.J., R.O., E.S., Z.H., H.Z., H.C., P.Y. and C.L. contributed to the analysis of all data and discussion. R.R. and E.S. performed and analyzed the STEM experiments. Y.C. and D.J. performed and analyzed the PFM experiments. X.T., P.Y., and R.O. performed and analyzed the XRD experiments. B.C. performed and analyzed the ellipsometry experiments. K.B. and C.-C.C. conducted the band simulation. Z.H. and D.J. wrote the original draft. All authors contributed to the paper and the interpretation of data.

#### Data availability

All data are available in the paper and Supplementary Information. Extra data are available from the corresponding authors upon reasonable request.

#### Competing interests:

The authors declare no competing interests.

Supplementary Information for:

# Compositional gradient engineering for enhanced ferroelectricity in ultrathin AlScN

Zekun Hu[1], Haiwen Zhang[1], Rajeev Kumar Rai[2], Yuhong Cao[1], Xiaolei Tong[1], Pedram Yousefian[1], Hyunmin Cho[1], Bongjun Choi[1], Chao-Chuan Chen[1], Yunfei He[1], Kefei Bao[1], Chloe Leblanc[1], Eric A. Stach[1,2], Roy Olsson[1], Deep Jariwala[1,*]

[1]Department of Electrical and Systems Engineering, University of Pennsylvania, Philadelphia, Pennsylvania 19104, United States of America

[2]Department of Materials Science and Engineering, University of Pennsylvania, Philadelphia, Pennsylvania 19104, United States of America

[*]Correspondence should be addressed: dmj@seas.upenn.edu (D. J.)

Table of Contents

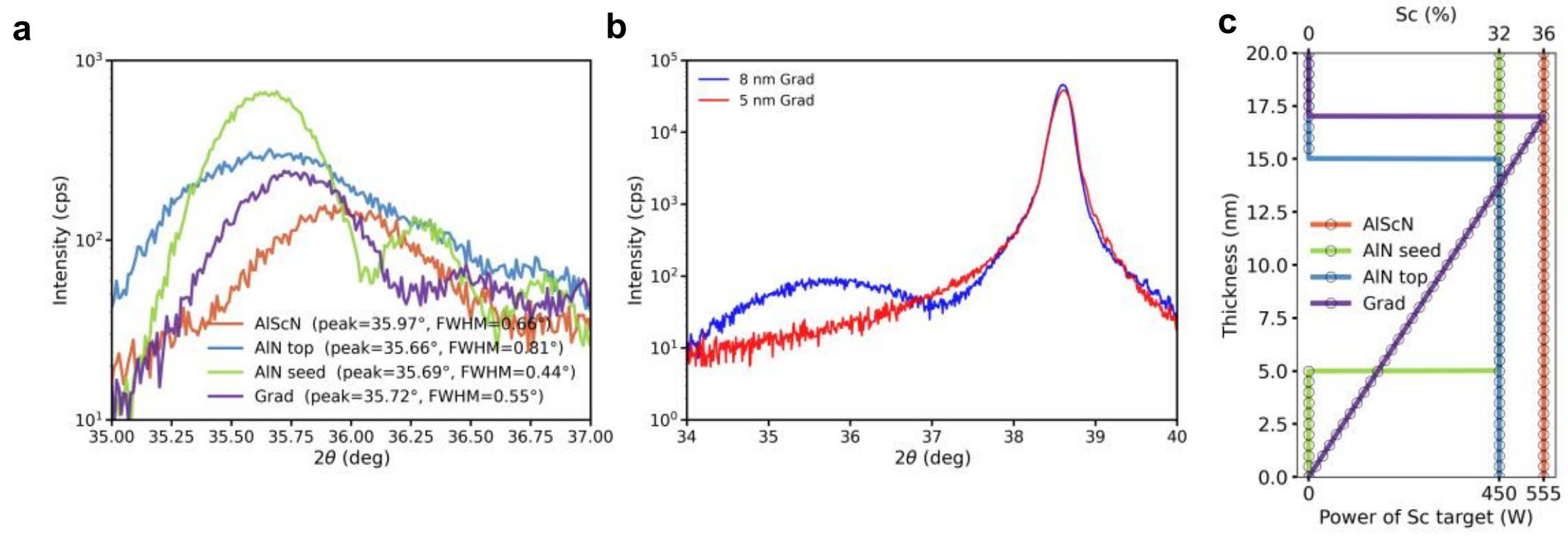

**Fig. S1. X-ray diffraction analysis of the AlN/AlScN films and concentration control.**

**a,** Enlarged view of the overlapping AlN/AlScN diffraction region for the four reference films, with the extracted peak position and full width at half maximum (FWHM) indicated for each sample. The AlN-seed film shows the smallest FWHM of the four. **b,** X-ray diffraction patterns of the thinner graded films. The 8 nm graded film retains a detectable diffraction feature, whereas the 5 nm graded film produces a signal that is difficult to resolve clearly. **c,** Programmed Sc target power and corresponding nominal Sc composition versus film thickness.

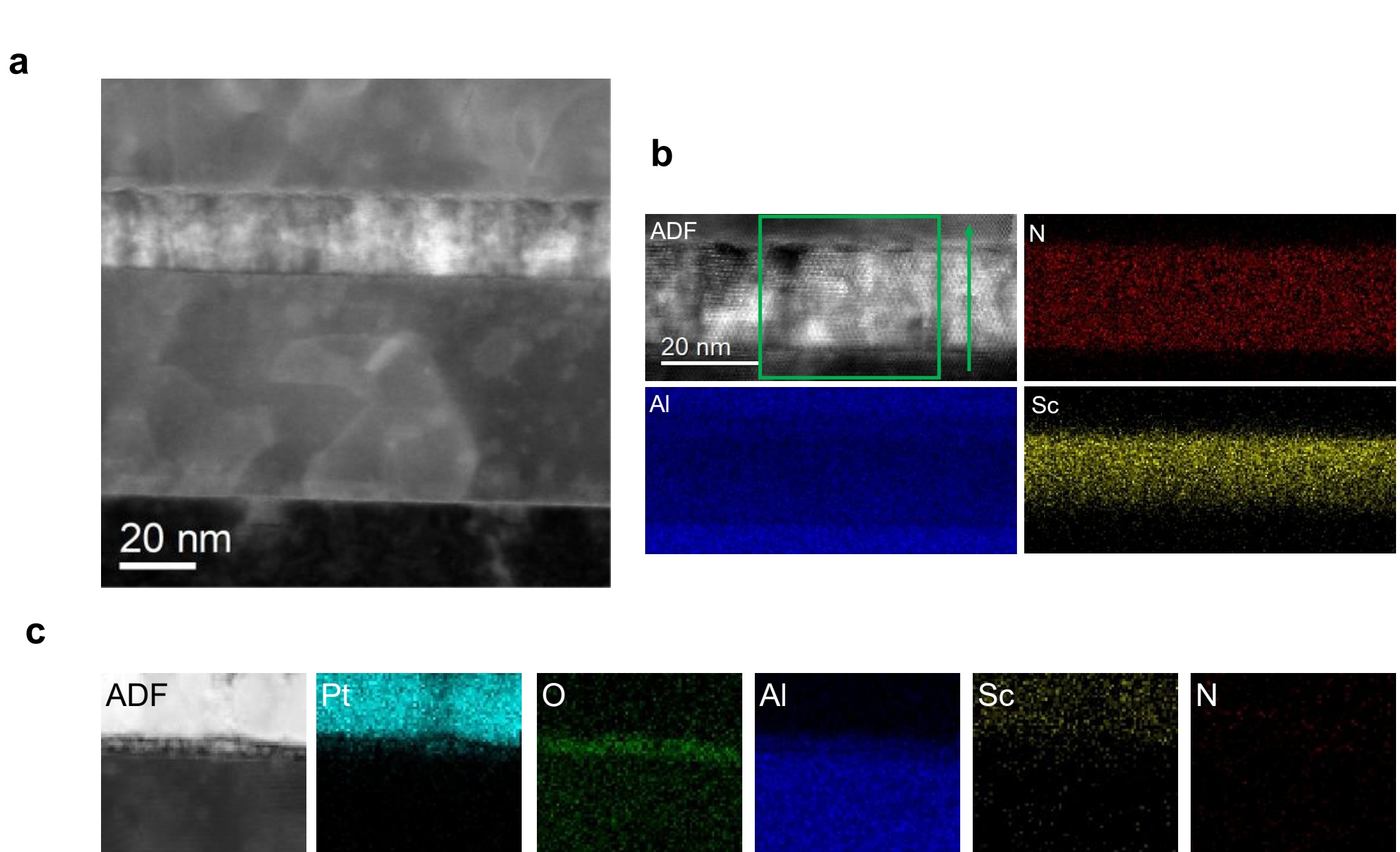

c

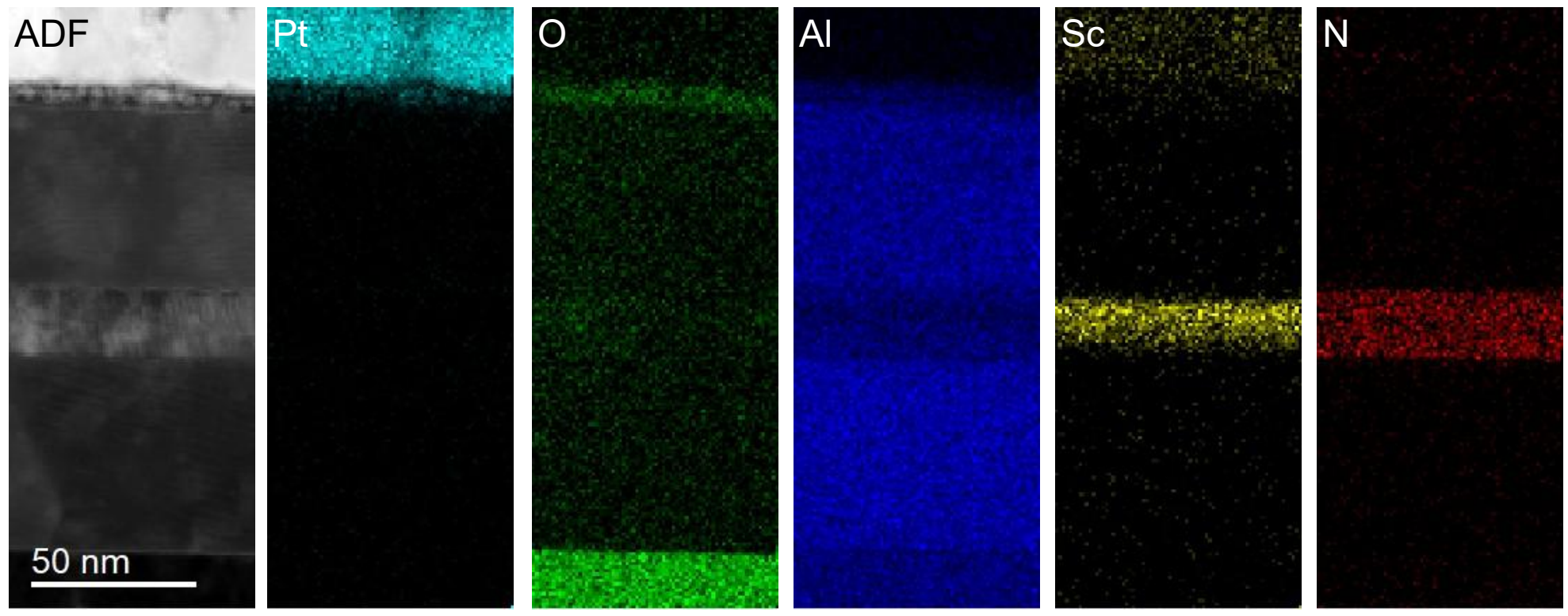

**Fig. S2. Cross-sectional STEM–EDS analysis of the 20 nm Al/AlScN/Al film stack.**

**a,** Low magnification ADF STEM image of the graded AlN–AlScN film, showing the continuous ferroelectric layer between the top and bottom electrodes. **b,** Zoomed ADF STEM image taken from the boxed region in **a**, showing the local microstructure of the graded heterostructure. The film retains continuous columnar contrast across the thickness without obvious structural discontinuity, consistent with the lattice matched compositional grading from AlN rich to AlScN rich regions. **c,** Cross sectional STEM EDS elemental maps of the same heterostructure, showing Pt, O, Al, Sc, and N distributions. The Sc signal is confined within the ferroelectric layer, while the Al and N maps confirm the nitride film continuity across the graded stack.

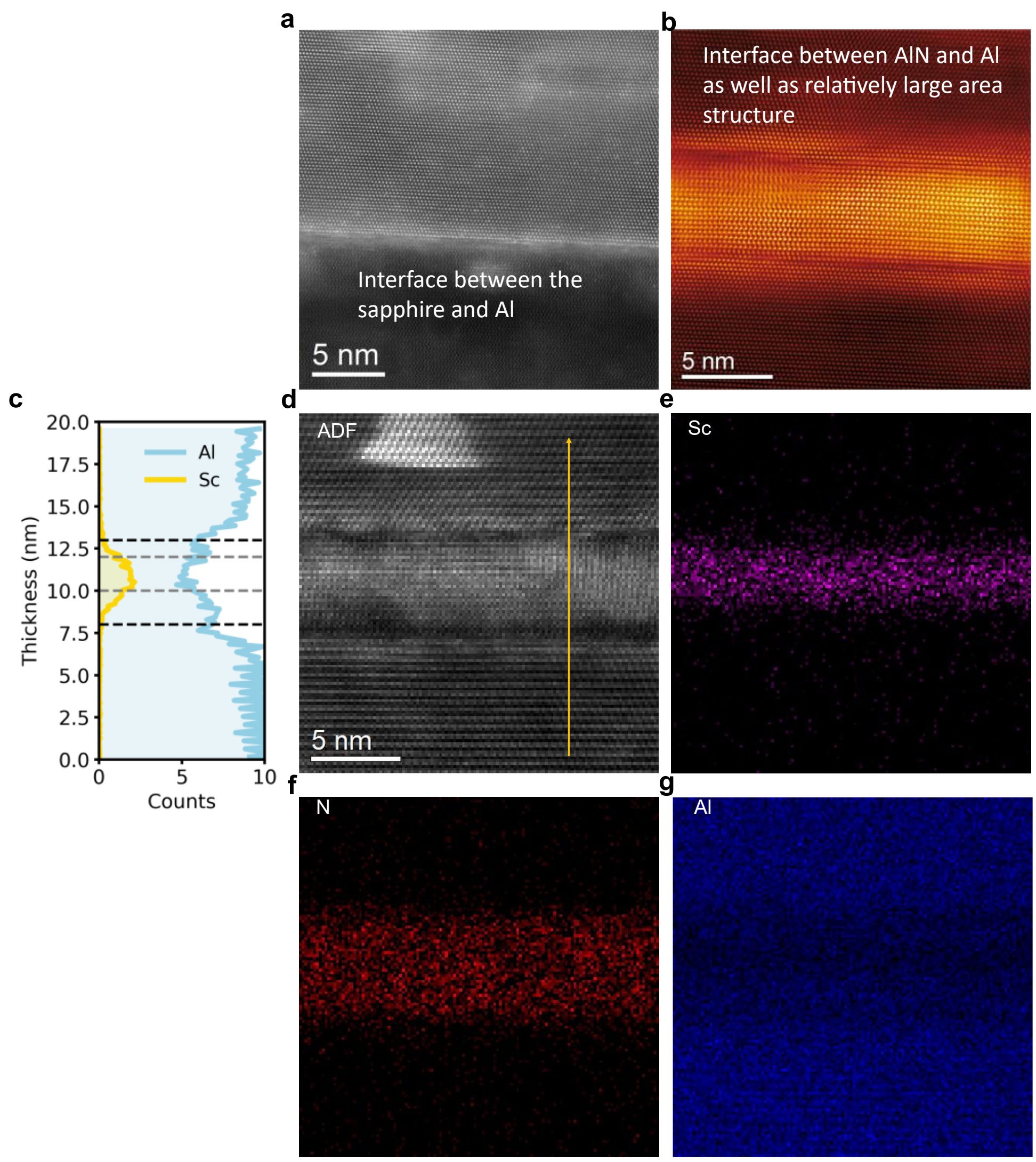

**Fig. S3. Atomic scale STEM and compositional analysis of the 5 nm graded AlN–AlScN heterostructure.**
**a,** Atomic resolution ADF STEM image showing the interface between the sapphire substrate and the epitaxial Al bottom electrode. **b,** Atomic resolution HAADF STEM image showing the AlN/Al interface and a relatively large area of ordered crystalline structure, confirming the high crystalline quality of the ultrathin stack. **c,** EDS line profile of Al and Sc counts across the film thickness, showing the confined Sc containing region within the graded heterostructure. The dashed lines mark the approximate boundaries of the designed ultrathin stack. **d,** Cross sectional ADF STEM image of the 5 nm graded AlN–AlScN heterostructure, showing a continuous ultrathin ferroelectric layer between the surrounding electrode regions. **e** to **g,** STEM EDS elemental maps of Sc, N, and Al, respectively, confirming Sc incorporation within the ultrathin ferroelectric layer and continuous Al and N distribution across the nitride stack.

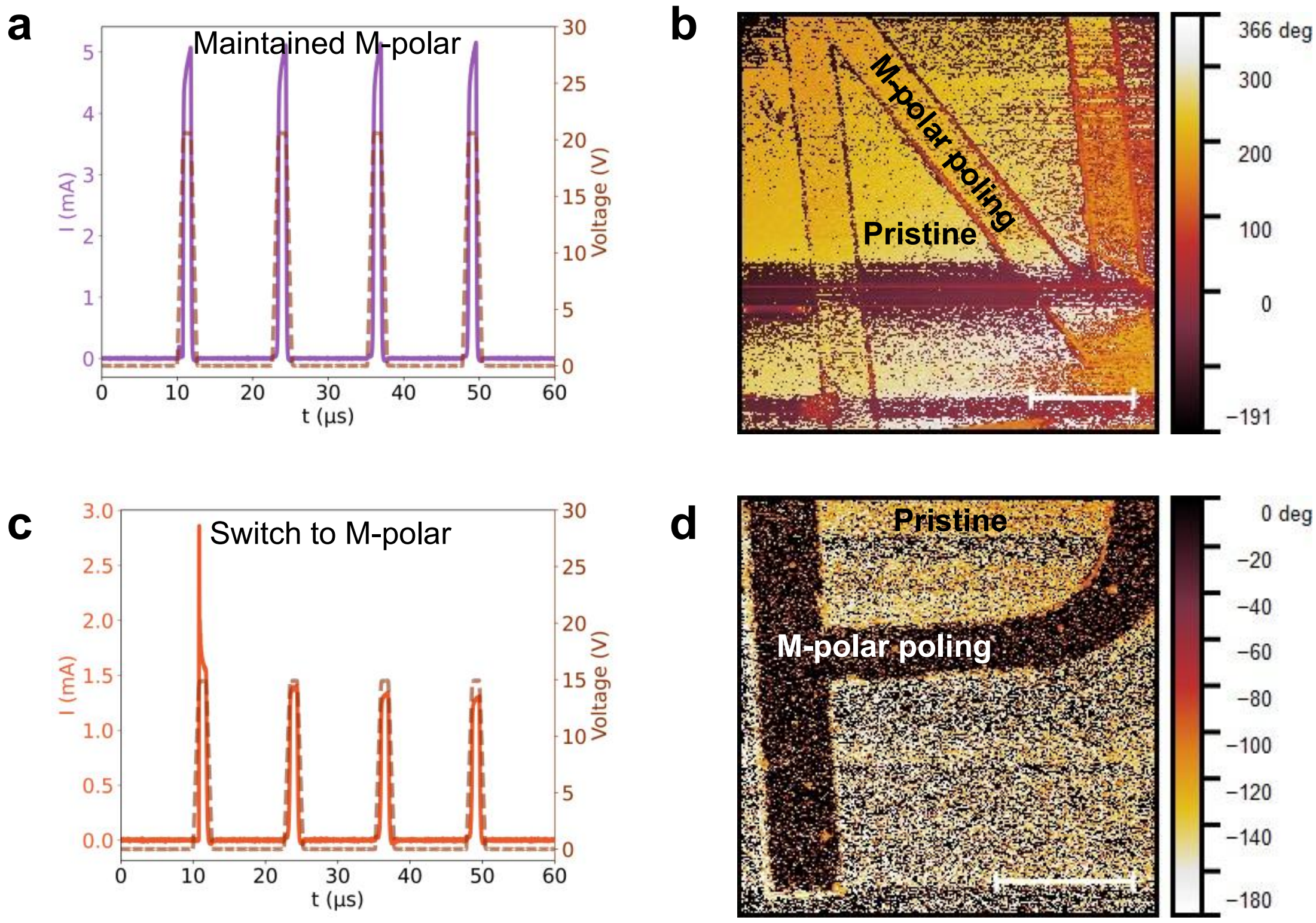

**Fig. S4. Comparison of metal poling behavior in graded and homogeneous AlScN films.**
**a,** Current response during the initial voltage pulse sequence applied to metal-pole the graded film. No switching current is observed, indicating that the polarization state remains unchanged under the applied pulses. **b,** PFM phase image after writing the letter "N" using metal poling on the graded film. The poled region shows no discernible phase contrast relative to the surrounding pristine area, confirming the absence of polarization switching. **c,** Current response during the initial voltage pulse sequence applied to metal-pole the homogeneous AlScN film. A pronounced switching current is observed at the beginning of the pulse sequence, indicating polarization reversal. **d,** PFM phase image after writing the letter "P" by metal poling on the AlScN film. The poled region exhibits an approximately 180° phase contrast compared with the surrounding pristine region, consistent with successful ferroelectric polarization switching.

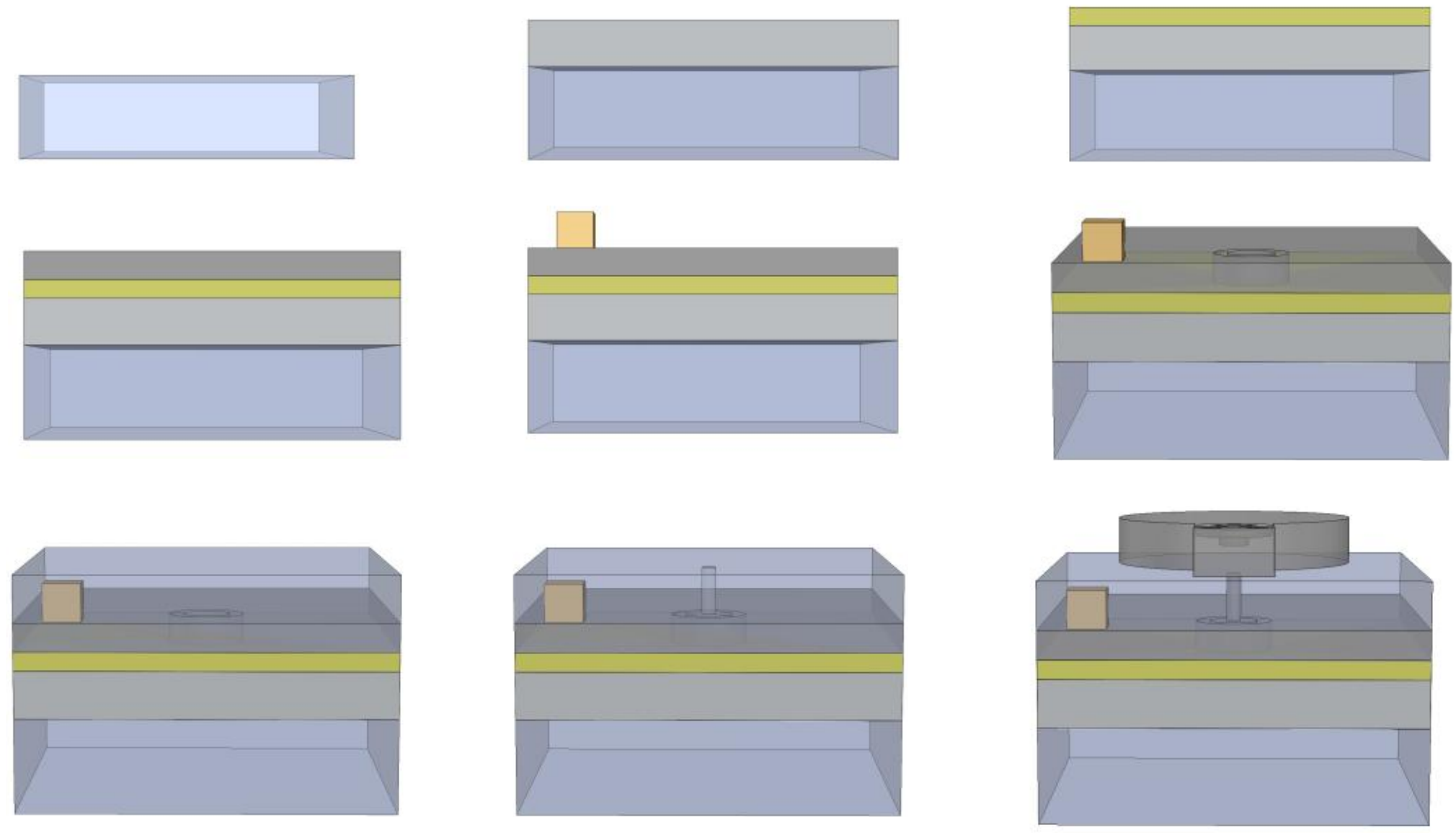

**Fig. S5. Schematic illustration of the device fabrication process.** Starting from an in situ grown Al/AlScN/Al stack on sapphire, Au alignment markers are first deposited. The top Al layer is then patterned by ICP etching to define the device area. A $SiO_2$ passivation layer is deposited, followed by via patterning and etching to expose the contact region. After sputter etching of the exposed surface, a homogeneous Al layer is deposited and subsequently patterned to form the contact pads.

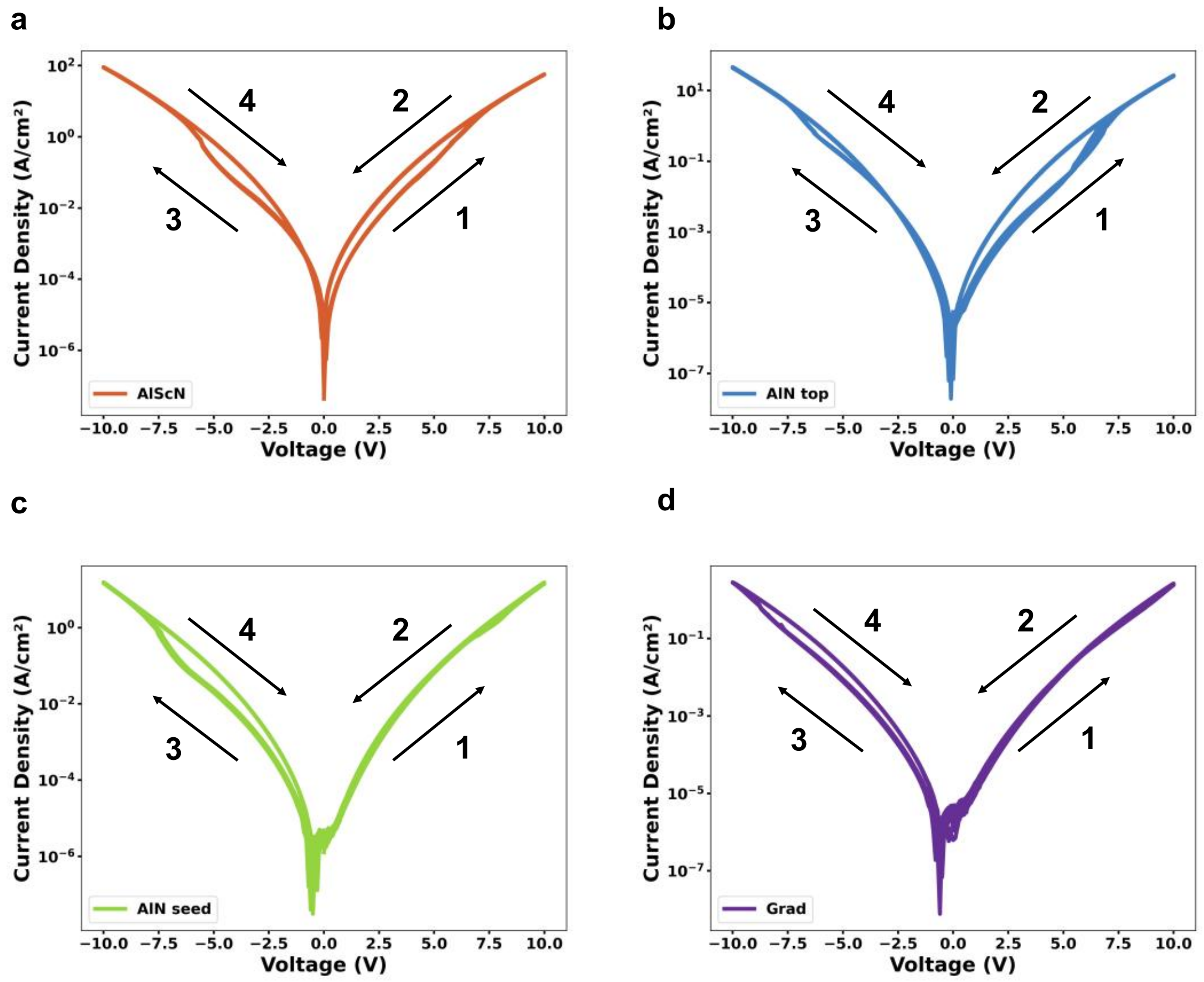

**Fig. S6. Device-to-device DC current density–voltage characteristics of the four stacks.**
DC current density–voltage (J–V) curves measured from n = 10 devices for each of the four stacks: **a,** AlScN, **b,** AlN-top, **c,** AlN-seed, and **d,** graded. The grouped traces show consistent device-to-device behavior within each stack, indicating good uniformity and reproducibility.

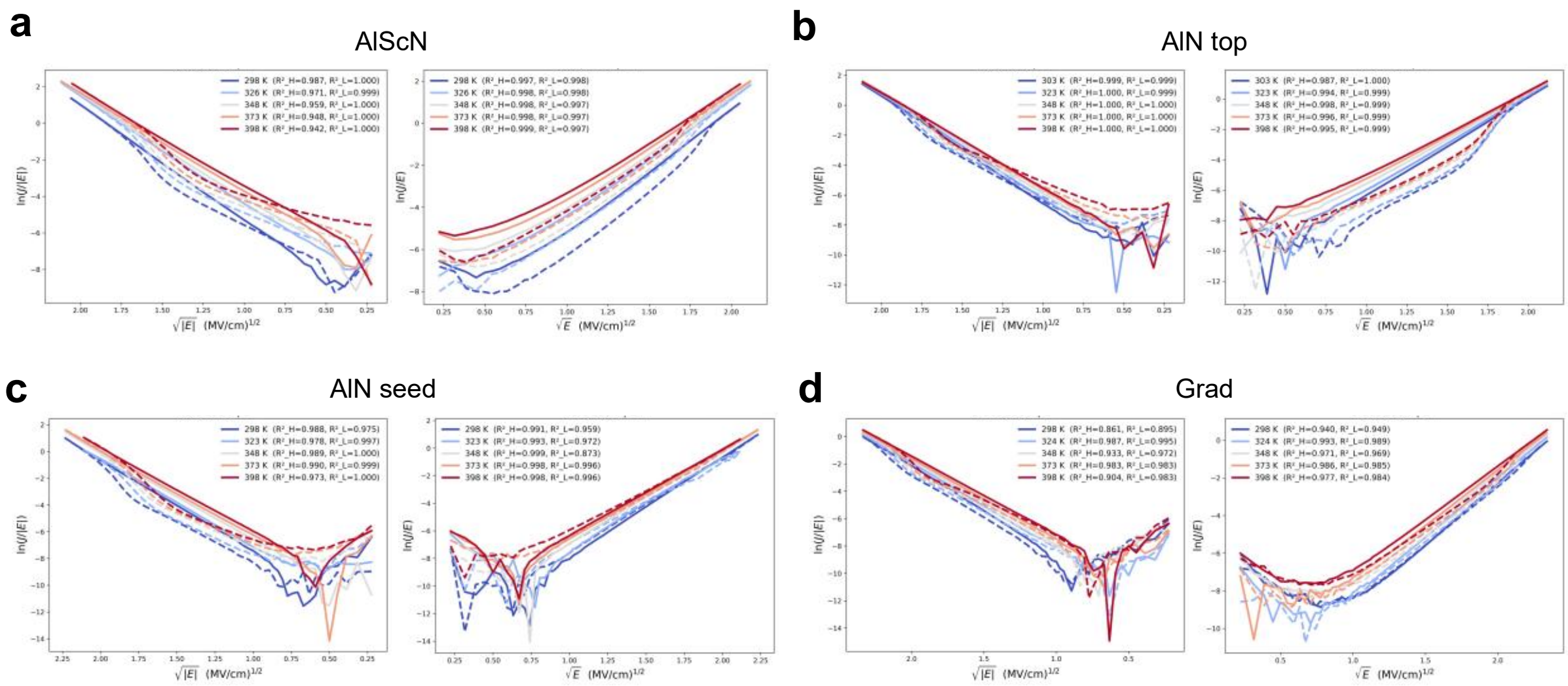

**Fig. S7. Poole–Frenkel fitting of leakage conduction in the four stacks.**

Temperature-dependent linearized leakage-current plots for the four stacks under both bias polarities, shown together with the corresponding linear fits. Solid lines denote the low-resistance state, while dashed lines denote the high-resistance state. The near-linear behavior in the Poole–Frenkel representation indicates that the leakage current is consistent with Poole–Frenkel-dominated conduction over the analyzed field range.

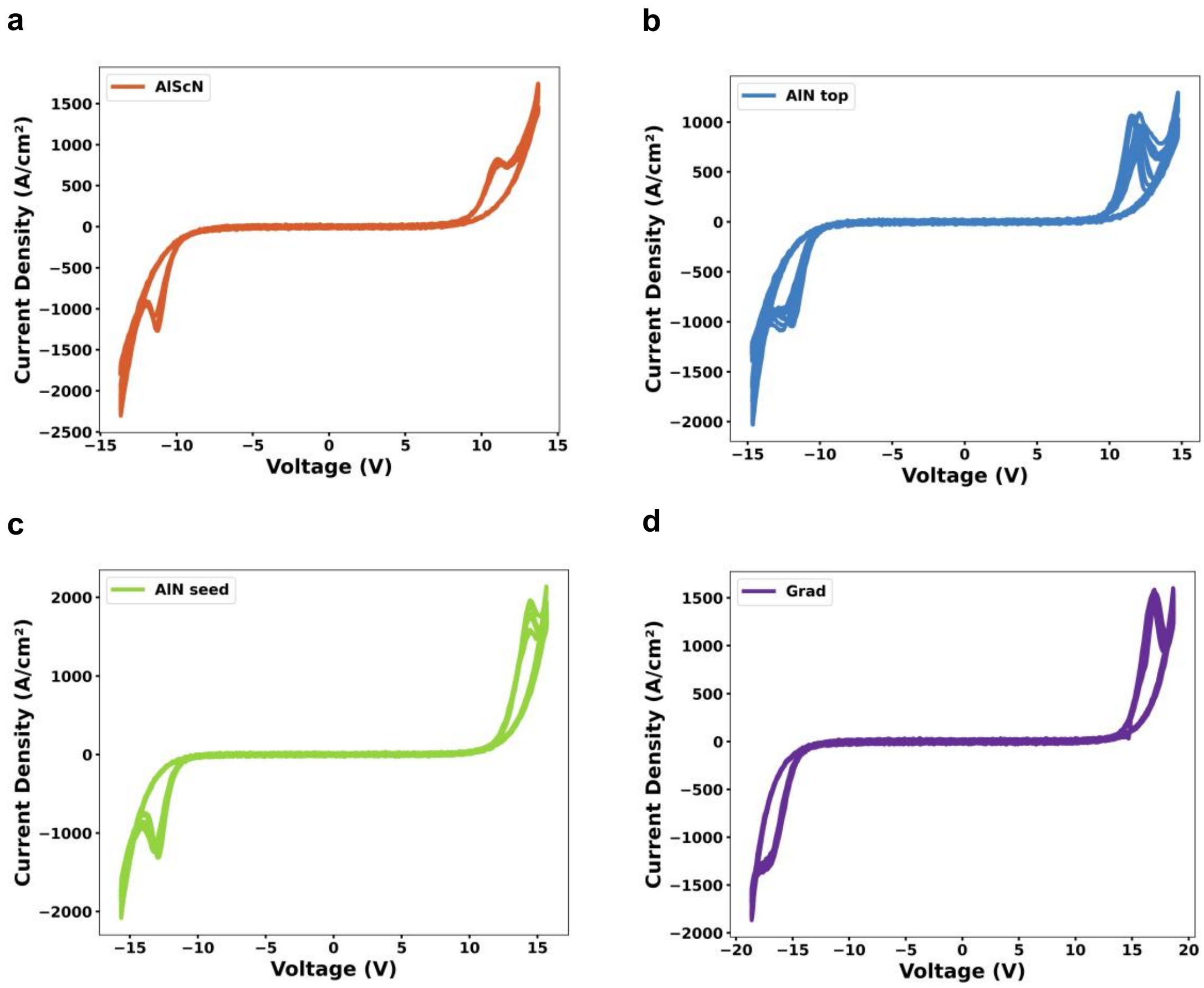

**Fig. S8. Device-to-device AC current density–voltage characteristics of the four stacks.**

AC current density–voltage curves measured from n=10 devices for each of the four stacks under 100 kHz: **a,** AlScN, **b,** AlN-top, **c,** AlN-seed, and **d,** Graded. The grouped traces show consistent device-to-device behavior within each stack, indicating good uniformity and reproducibility.

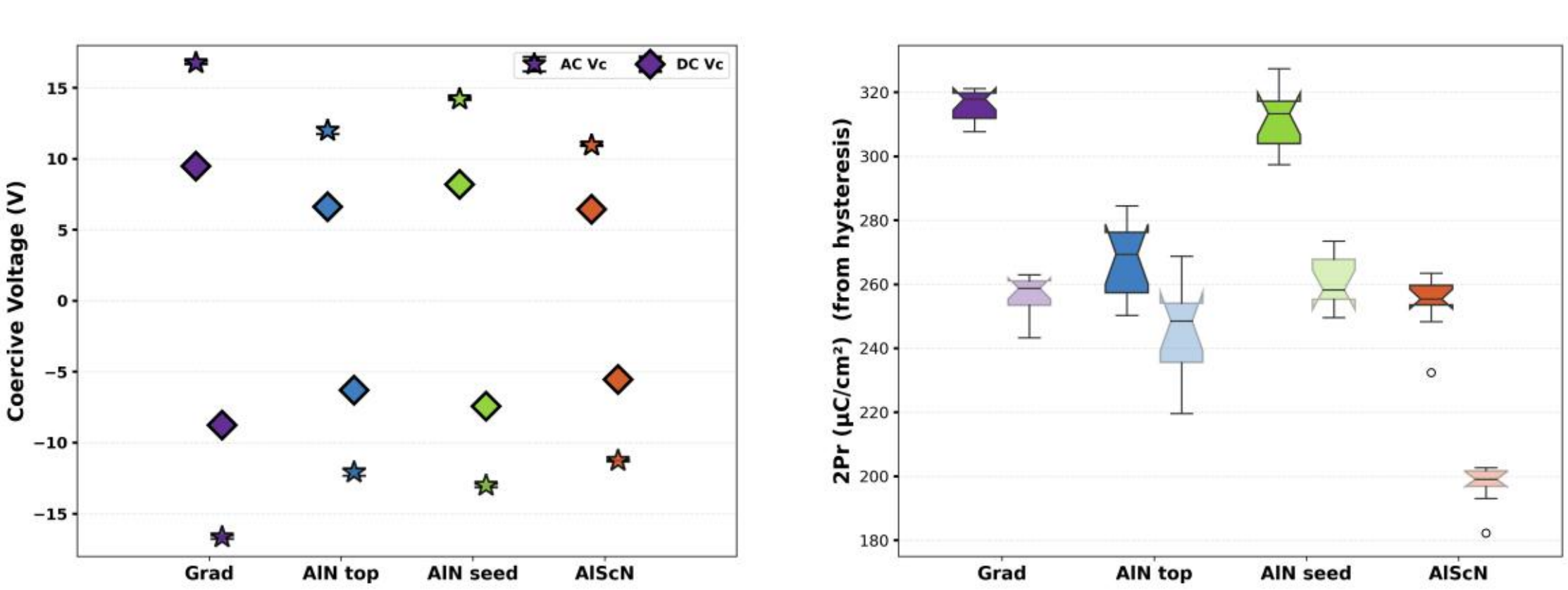

**Fig. S9. Summary of switching voltage and polarization extracted from AC measurements.**
**a,** Comparison of the coercive voltage extracted from AC and DC measurements for the four stacks. **b,** Box plots summarizing the 2Pr values extracted from the hysteresis loops for the four stacks, measured across $n = 10$devices per stack.

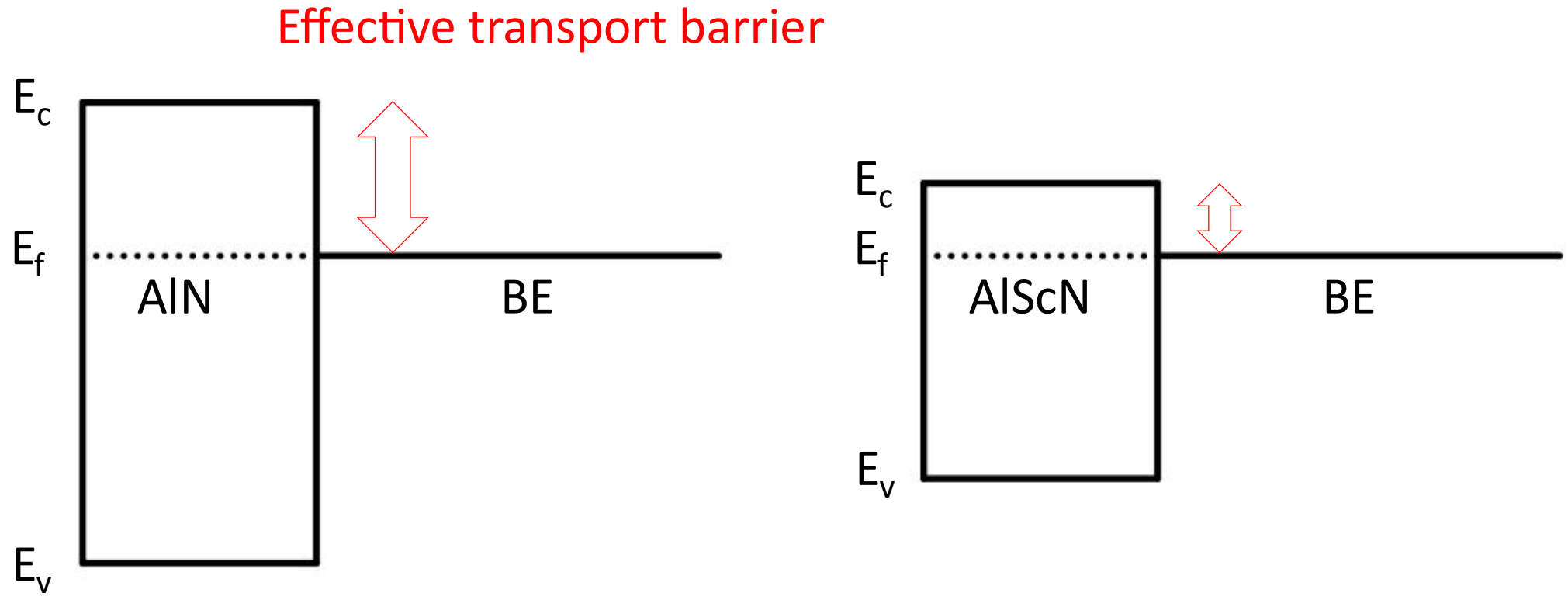

**Fig. S10. Schematic illustration of tunnelling barrier modulation by the AlN interface.**
Band schematics comparing the effective transport barrier height for AlN and AlScN at the electrode interface. The AlN layer produces a larger effective tunnelling barrier, consistent with reduced charge injection across the interface.

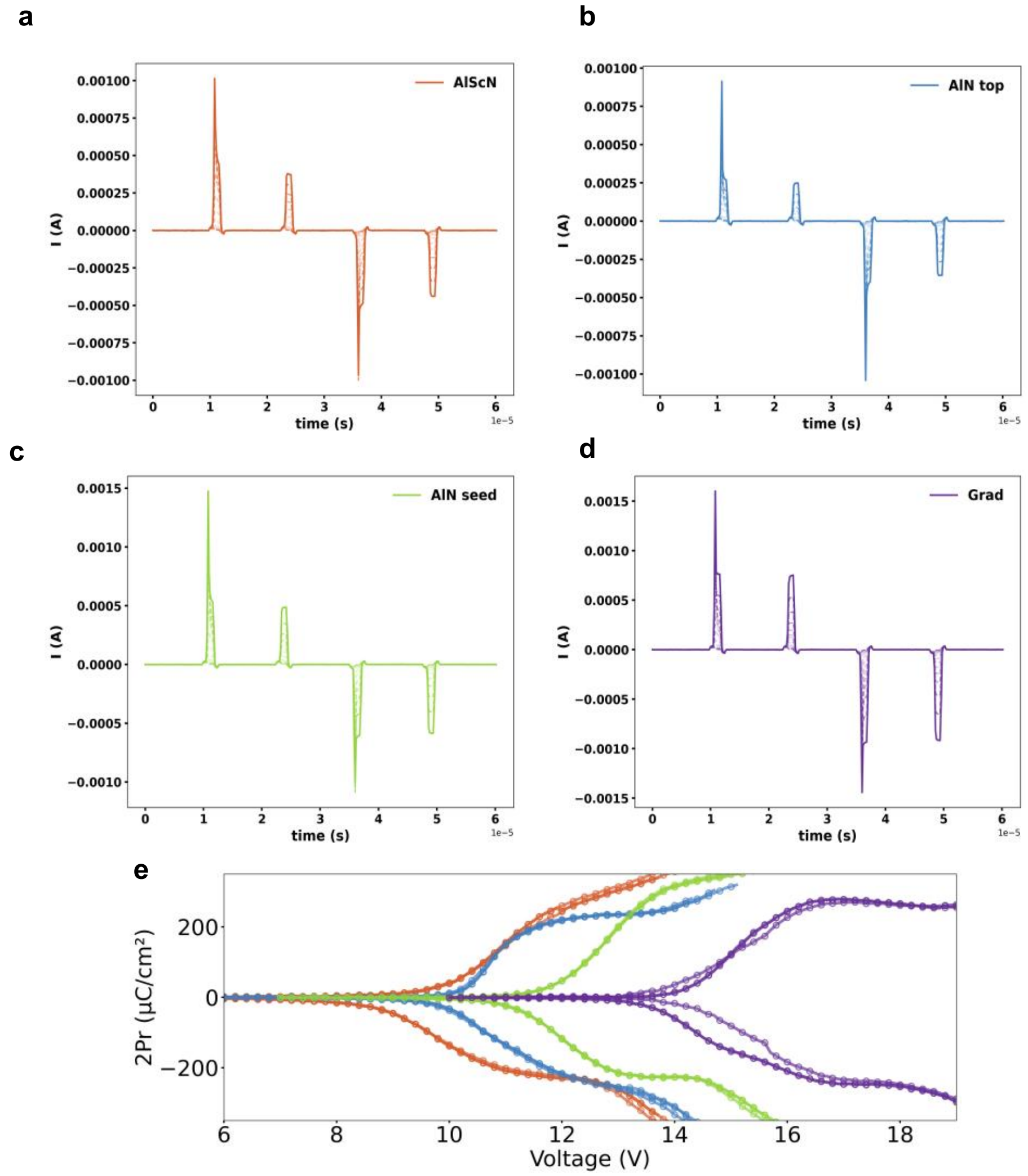

**Fig. S11. Time-domain PUND responses under gradually increasing pulse voltage.**

Detailed PUND current transients for the four stacks, **a,** AlScN, **b,** AlN-top, **c,** AlN-seed, and **d,** Graded, showing the progressive emergence of switching with increasing pulse amplitude. **e,** PUND measurements with 0.1 V step resolution of 3 devices from each film.

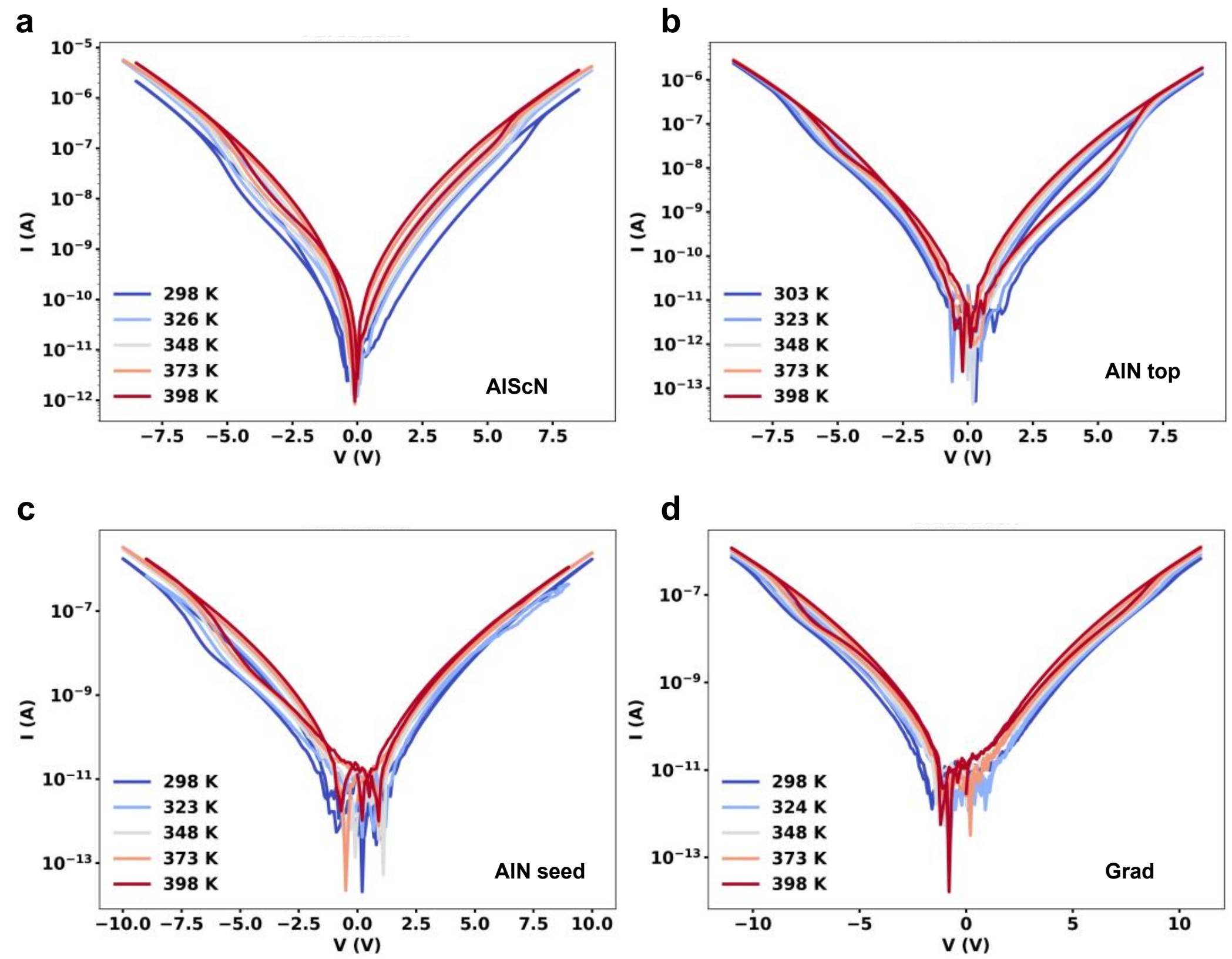

**Fig. S12. Temperature-dependent DC current–voltage characteristics of the four stacks.**

**a–d,** DC current–voltage curves measured over 298–398 K for **a,** homogeneous AlScN, **b,** AlN-top, **c,** AlN-seed and **d,** Graded. The temperature dependence of the leakage current is shown for each structure.

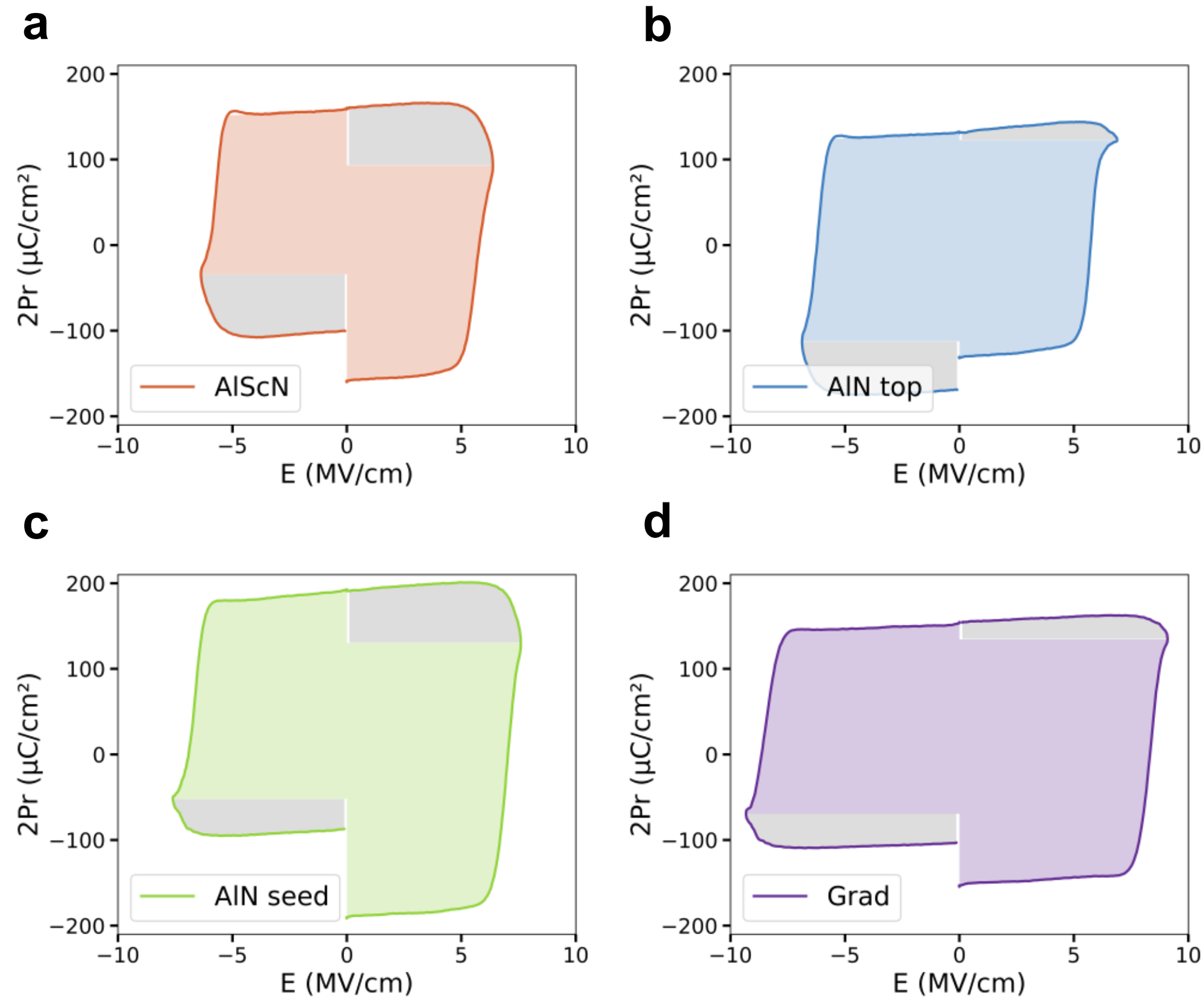

**Fig. S13. Triangular PUND loops with leakage subtraction.**
**a–d,** Triangular PUND responses for **a,** AlScN, **b,** AlN-top, **c,** AlN-seed and d, Graded stacks. Grey shaded regions denote the leakage-induced charge contribution. Relative to homogeneous AlScN, the leakage contribution is smaller on the negative-bias side for AlN-seed, smaller on the positive-bias side for AlN-top, and reduced on both sides for the graded stack with AlN interfaces on both sides.

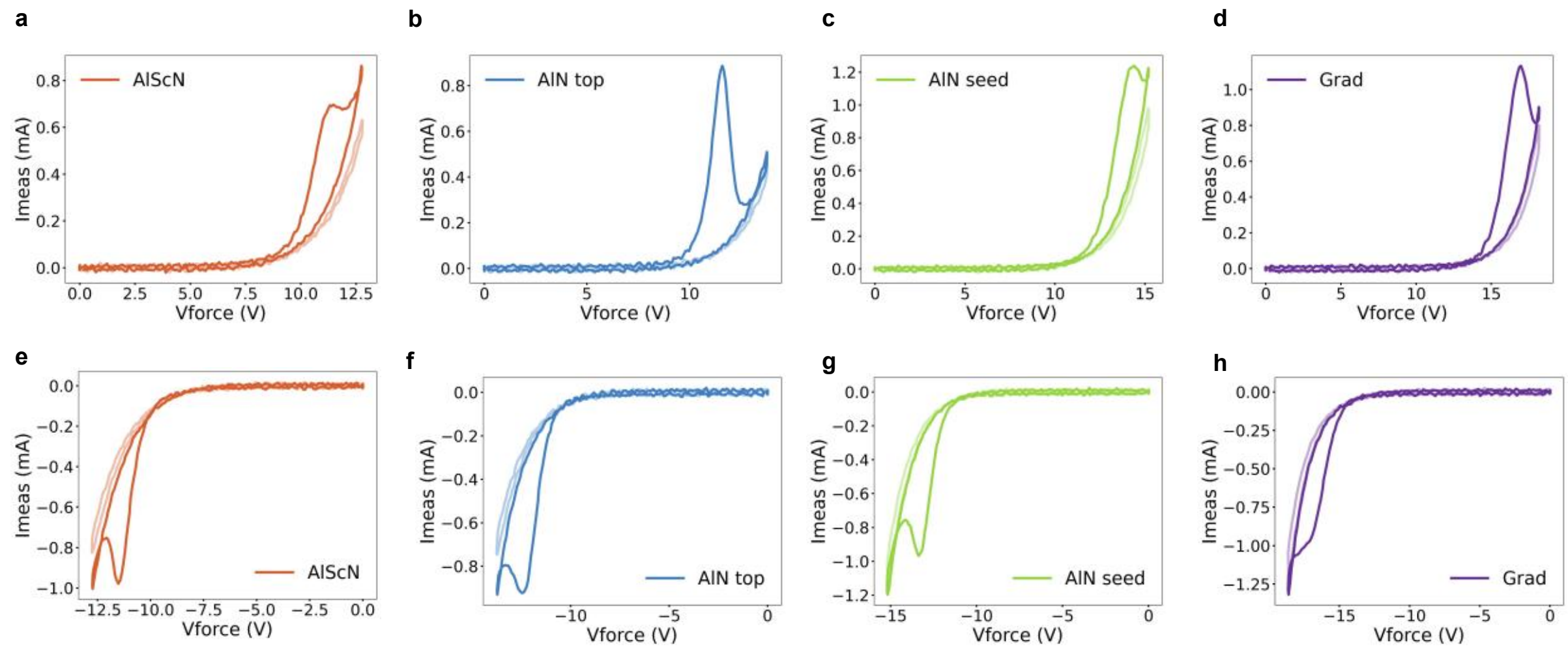

**Fig. S14. Current response from triangular PUND measurements.**
Current–voltage responses for the **a, e,** AlScN, **b, f,** AlN-top, **c, g,** AlN-seed and **d, h,** Graded extracted from triangular PUND measurements under positive and negative bias. The separation between the backward sweep and the second loop is consistent with post-switching leakage. This mismatch is reduced on the positive-bias side for AlN-top, on the negative-bias side for AlN-seed, and on both sides for the graded stack, consistent with the presence of AlN interfacial layers.

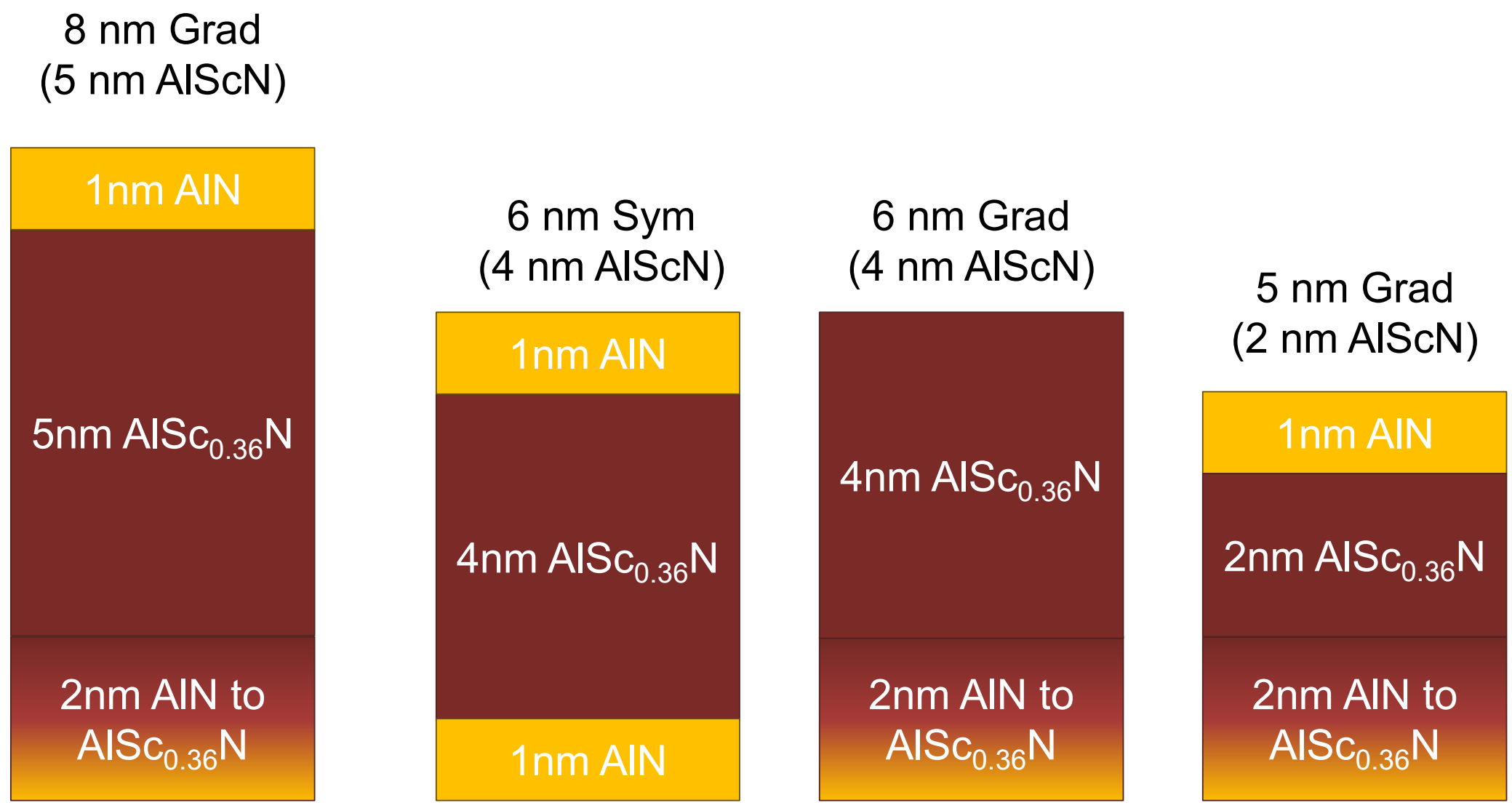

**Fig. S15. Schematic structures of the ultrathin stacks.**
Schematics of the ultrathin graded and reference film designs, showing the layer configuration and nominal thickness of each structure: 8 nm graded, 6 nm symmetric AlN, 6 nm graded and 5 nm graded.

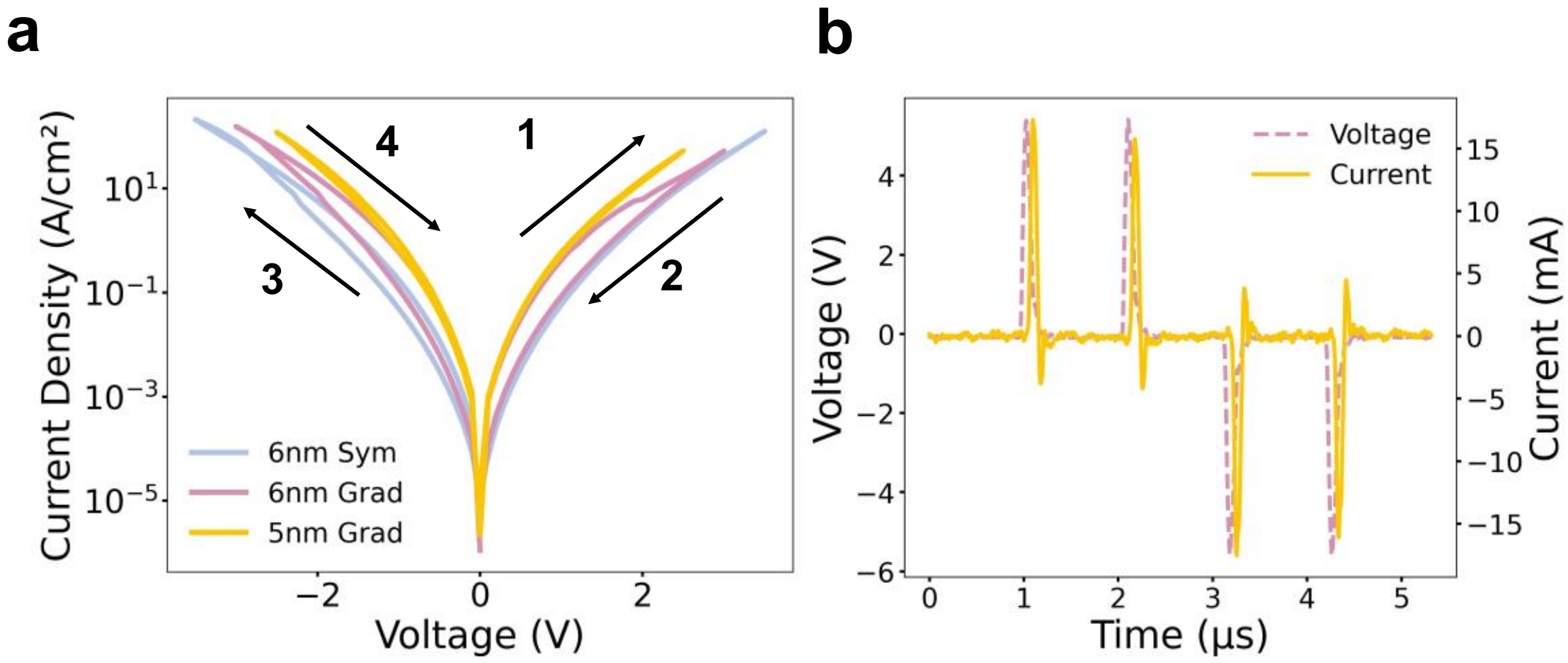

**Fig. S16. Electrical response of ultrathin films from DC to 50 ns pulse measurements.**

a, DC current density–voltage characteristics of the 6 nm symmetric, 6 nm graded and 5 nm graded stacks, with a switching-related feature observable under negative bias. b, Representative pulsed voltage and current responses of the 5 nm Grad devices with 5 µm radius, measured using voltage pulses with a 30 ns rise/fall time and a 20 ns pulse hold time. The difference in peak height between the P and U responses, and between the N and D responses, is consistent with polarization switching in the ultrathin films.

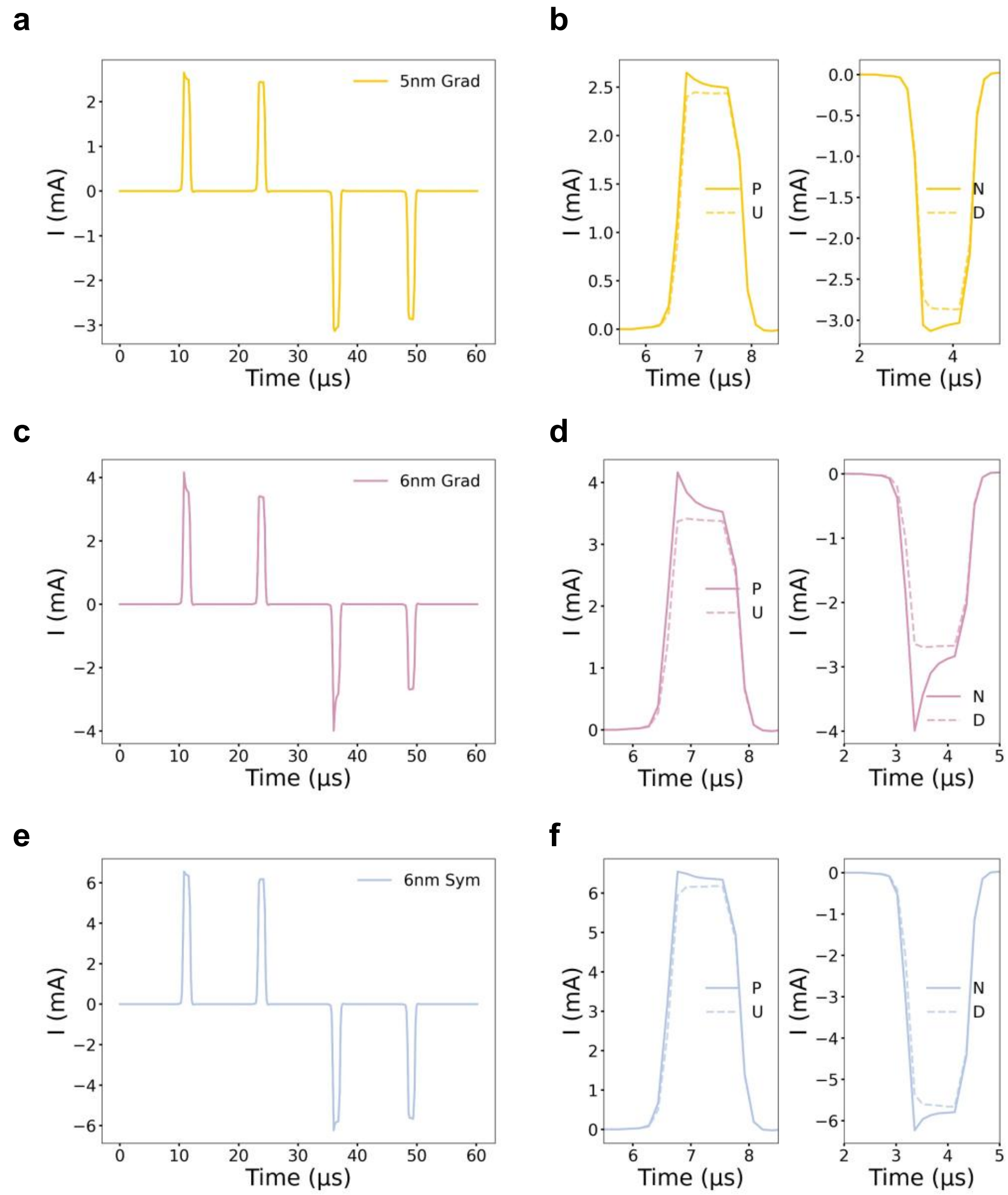

**Fig. S17. Detailed PUND responses of the ultrathin stacks.**
**a, c, e,** Time-domain PUND current traces for the **a,** 5 nm graded, **c,** 6 nm graded and **e,** 6 nm symmetric films.
**b, d, f,** Enlarged views of the corresponding P/U and N/D responses for **b,** 5 nm graded, **d,** 6 nm graded and **f,** 6 nm symmetric films. The difference between P and U, and between N and D, resolves the switching contribution and is consistent with polarization switching in each ultrathin stack.

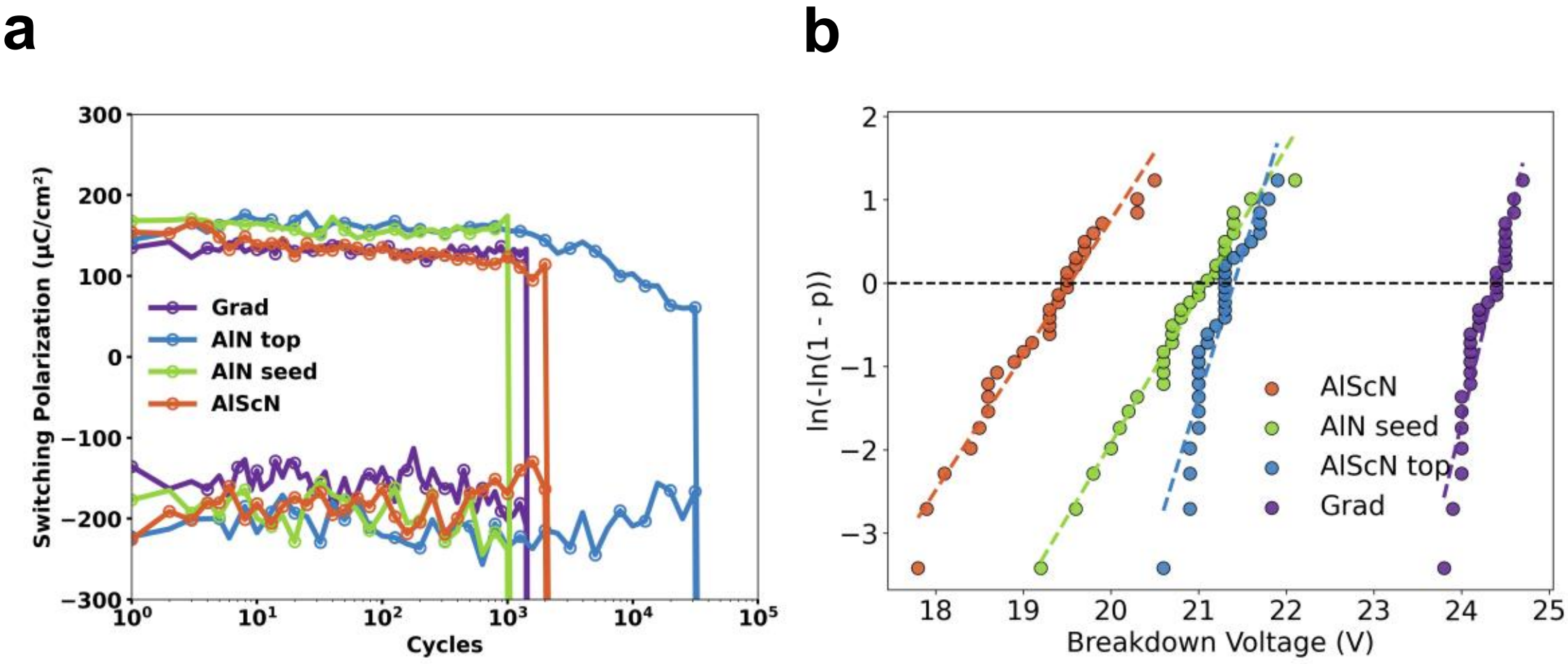

**Fig. S18. Endurance and Weibull analysis of the four 20 nm stacks.**
**a,** Switching polarization as a function of PUND cycling number for the graded, AlN-top, AlN-seed and homogeneous AlScN stacks. The AlN-top stack shows comparatively extended endurance, whereas the other three stacks exhibit similar cycling behavior. **b,** Weibull plots of breakdown voltage for the four stacks. The graded stacks show relatively tighter distributions together with higher breakdown voltages, indicating improved uniformity and dielectric robustness.

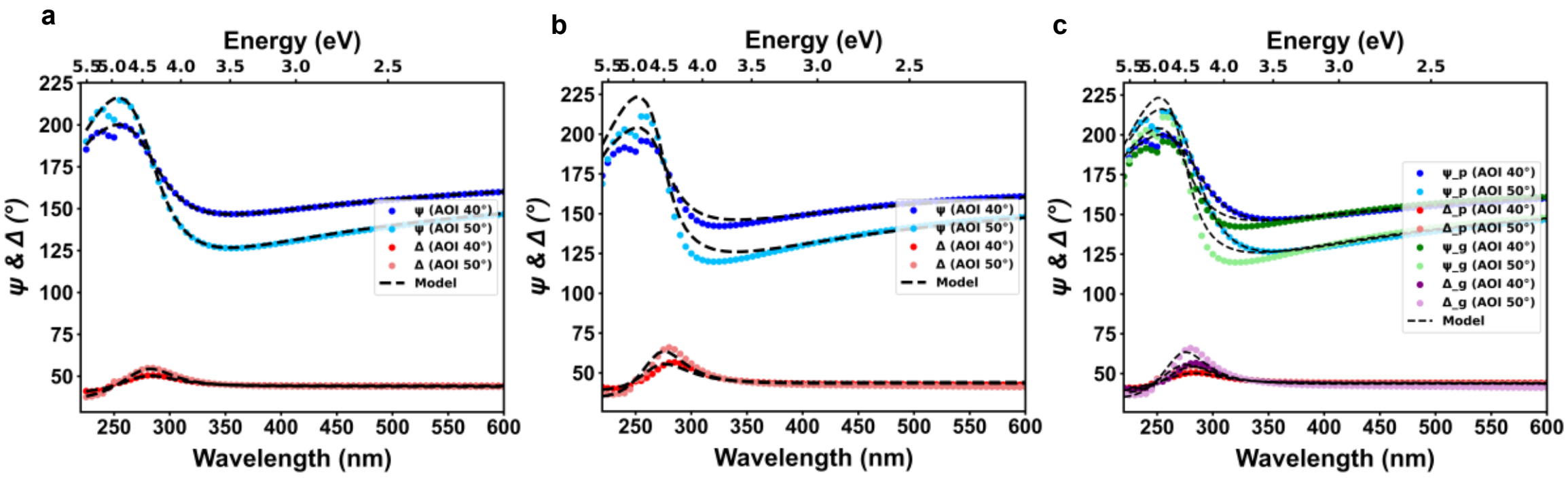

**Fig. S19. UV-range spectroscopic ellipsometry of homogeneous and graded films.**
**a,** Angle-dependent ellipsometry spectra (Ψ and Δ) of the 20 nm homogeneous AlScN film together with the fitted model (RMSE ~1). **b,** Angle-dependent ellipsometry spectra of the 20 nm graded film together with the fitted model (RMSE ~4). **c,** Comparison of the raw spectra for the two films, showing clearly distinct optical responses.

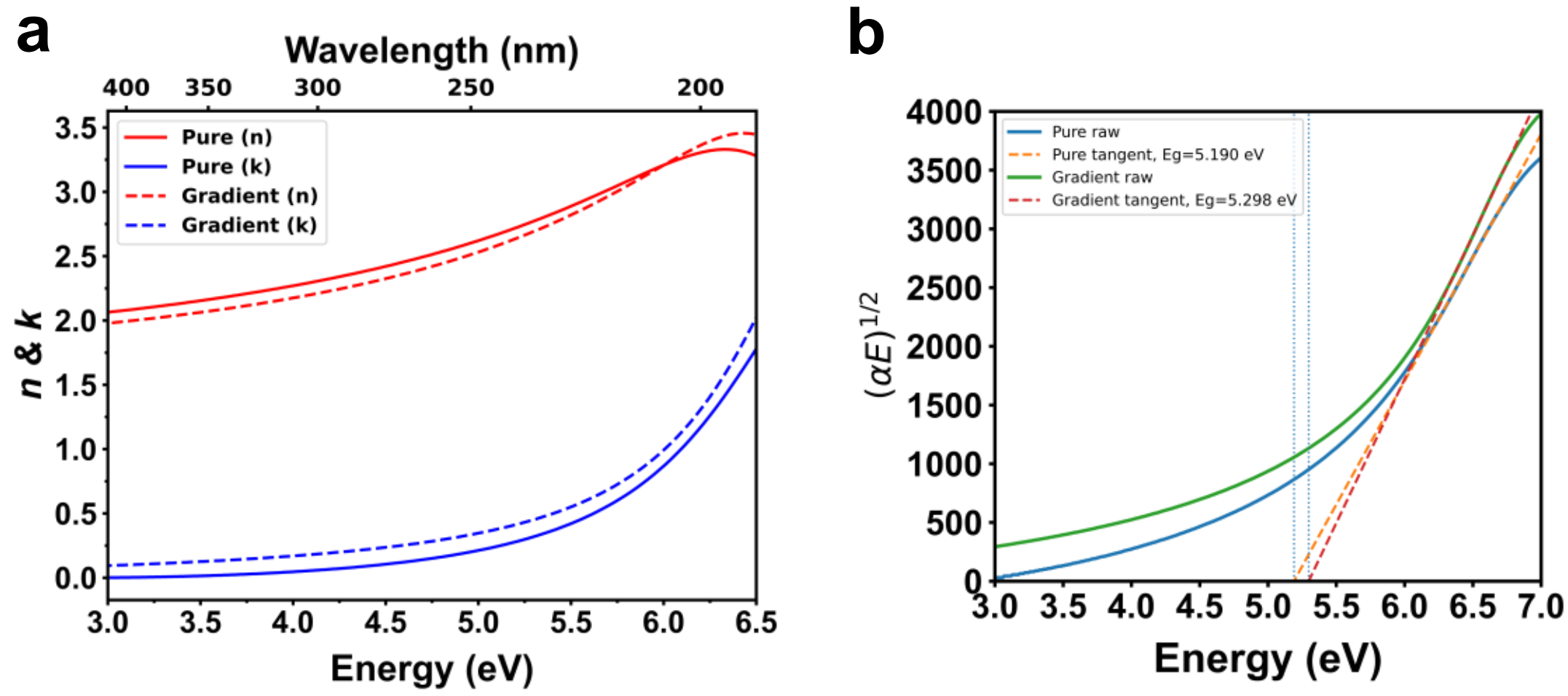

**Fig. S20. Optical constants and indirect bandgap fitting from ellipsometry.**
**a,** Refractive index (n) and extinction coefficient (k) extracted from ellipsometry for the homogeneous AlScN and graded films. **b,** Tauc plots with linear fitting used to estimate the indirect optical bandgap of each film.

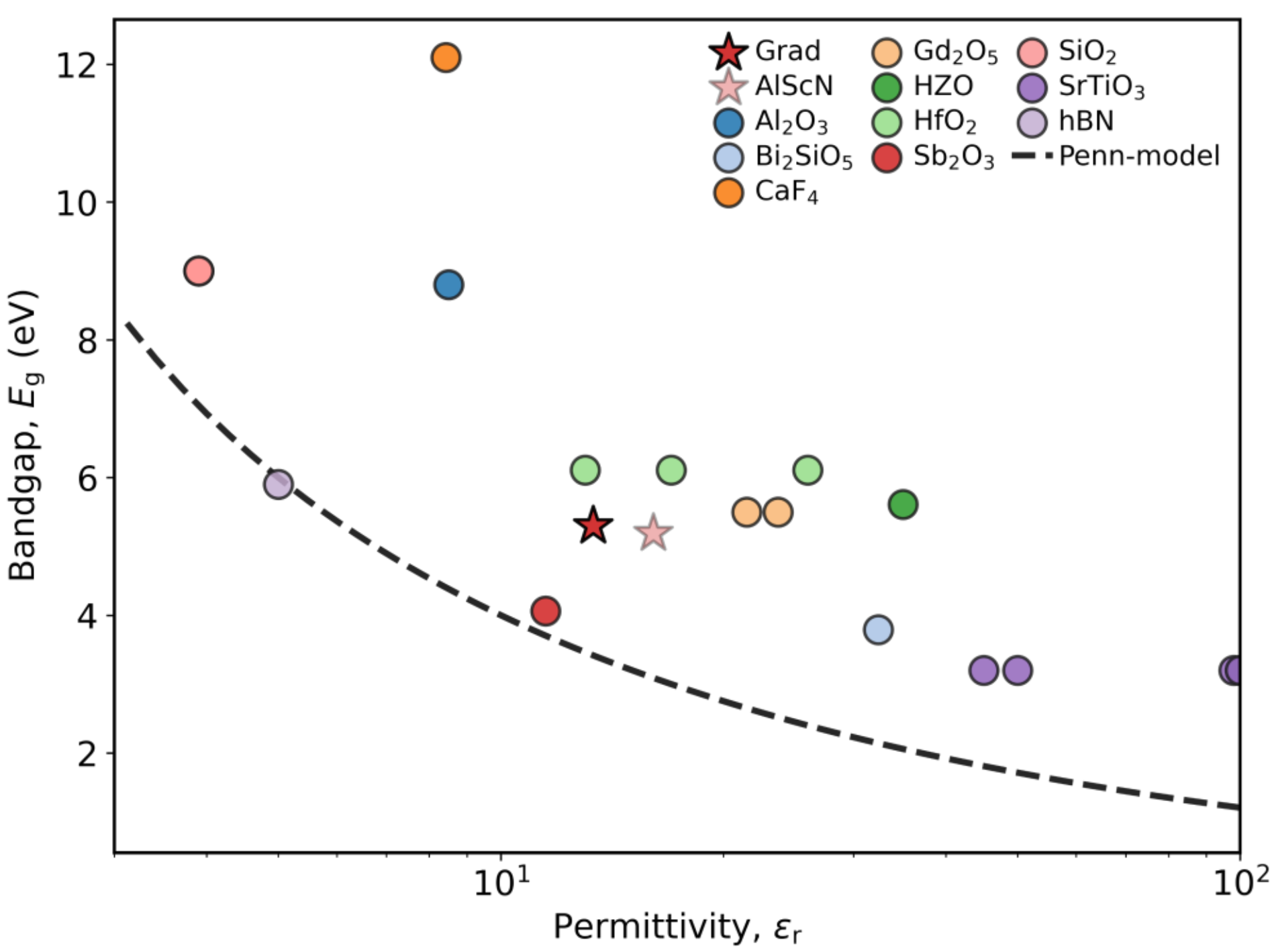

**Fig. S21. Bandgap benchmark.**

Bandgap, $E_g$, plotted as a function of relative permittivity, $\varepsilon_r$, for selected dielectric and ferroelectric materials. The graded AlScN heterostructure and reference AlScN sample are highlighted with star markers. The dashed line shows a Penn-model-type trend[1], illustrating the general inverse correlation between dielectric constant and bandgap. The bandgap values of the graded AlScN heterostructure and AlScN reference are optically measured using Tauc-method fitting, whereas the Eg values of the other materials[2, 3, 4, 5, 6, 7, 8, 9, 10] are mainly taken from DFT-reported literature values.

**Table S1. As-grown polarization state of the AlScN films.**

| Growth Method | Material | Thickness (nm) | Substrate | Polarity |
|---|---|---|---|---|
| MOCVD[11] | $Al_{0.85}Sc_{0.15}N$ | 230 | M-polar GaN (epitaxial) | M-polar |
| MBE[12] | $Al_{0.82}Sc_{0.18}N$ | 80 | M-polar GaN (epitaxial) | M-polar |
| MBE[13] | $Al_{0.82}Sc_{0.18}N$ | 80-100 | 4H-SiC / GaN (epitaxial) | N-polar |
| RF Sputter[14, 15] | (Power difference) | ~100-550 | Si(111), ITO/YSZ (epitaxial buffer) | M-polar/ N-polar |
| RF Sputter[14, 15, 16] | $Al_{1-x}Sc_xN$ ($x<0.12$) | ~100-550 | Si(111), Pt/Mo/YSZ (non-epitaxial or textured) | M-polar |
| | $Al_{1-x}Sc_xN$ ($0.16<x<0.36$) | | | N-polar |
| DC Sputter[17] | AlScN | 320 | AlN / Si (textured) | M-polar |
| DC Sputter[17] | AlScN | 480-1000 | AlN / Si (textured) | N-polar |
| DC sputter (this work) | AlScN | 20 | epitaxial Al | N-polar |
| DC sputter (this work) | Graded | 20 | epitaxial Al | M-polar |
| DC sputter (this work) | Graded | 5 | epitaxial Al | M-polar |

**Table S2. Benchmark of ultrathin ferroelectric films.**

| Material (ref.) | Thickness | Remanent Polarization | Coercive Field | Growth Method |
|---|---|---|---|---|
| CIPS[18] | - | 4 μC/cm² (200 nm) | 300 kV/cm | ME |
| HZO[19] | - | 16 μC/cm² (7.5 nm) | 1 MV/cm (7.5 nm) | ALD |
| HZO[20] | 5 nm | 34 μC/cm² (5 nm) | 5 MV/cm (5 nm) | PLD |
| HZO[21] | 1 nm | - | 2 MV/cm (1 nm) | ALD |
| PZTO[22] | 5 nm | 11 μC/cm² (5 nm)<br>150 μC/cm² (15 nm) | 1800 kV/cm (8 nm)<br>150 kV/cm (30 nm) | PLD |
| BTO[23] | 5 nm | 15 μC/cm² (5 nm)<br>36 μC/cm² (30 nm) | - | PLD |
| BFO[24] | 5 nm | 1 μC/cm² (10 nm) | ~2000 kV/cm (10 nm) | ALD |
| $TiO_2$[25] | < 3 nm | - | - | ALD |
| β′-$In_2Se_3$[26] | 5 Atomic Layer | 0.199 μC/cm² (DFT) | 0.27 eV/unit cell (DFT) | ME |
| SBTO[27] | 25 nm | 5 μC/cm² (25 nm) | 100 kV/cm (25 nm) | CVD |
| $Al_{1-x}Sc_xN$ (x=0.26) [28] | 3 nm | 100 μC/cm² (4 nm) | 10 MV/cm (4 nm) | RF Sputtering |
| $Al_{1-x}Sc_xN$ (x=0.3) [29] | < 5 nm | 80–100 μC/cm² (>400 nm) | 2–5 MV/cm (>400 nm) | RF Sputtering |
| $Al_{1-x}Sc_xN$ (graded) (this work) | 5 nm | > 100 μC/cm² | 6-8 MV/cm (10 kHz) | Sputtering |

**Note S1. Optical bandgap of homogeneous AlScN and graded films measured by UV spectroscopic ellipsometry.**

To estimate the bandgap of homogeneous $Al_{0.64}Sc_{0.36}N$ and gradient AlScN, the Tauc method was employed. The refractive indices of both samples were first extracted via spectroscopic ellipsometry, which provides high precision in determining optical constants. The measured ellipsometry parameters, ψ and Δ, exhibit a clear distinction between the two samples despite their structural similarity, reflecting the difference in refractive indices arising from the compositional variation (**Fig. S19**, ellipsometry fitting results). The absorption coefficient, α, was subsequently derived using the relation $\alpha = 4\pi k/\lambda$, where k is the extinction coefficient and λ is the wavelength. Tauc plot analysis was then applied according to the relation $(\alpha h\nu)^{1/n} = A(h\nu - E_g)$[30], where hν is the photon energy, $E_g$ is the optical bandgap, and A is a proportionality constant. Given that $Al_{0.64}Sc_{0.36}N$ exhibits an indirect bandgap character[31], an exponent of n = 2 was used for the fitting. From this analysis, bandgap values of 5.2 eV and 5.3 eV were extracted for homogeneous $Al_{0.64}Sc_{0.36}N$ and gradient AlScN, respectively (**Fig. S20**).

**Note S2. Write energy consumption calculation.**

The write energy of the 5 nm graded AlScN device was estimated from the high frequency PUND voltage and current waveforms shown in **Supplementary Fig. S16**. For each pulse, the pulse energy can be calculated by integrating the instantaneous electrical power over the pulse duration:

$$E_{write} = \int VIdt$$

E_write = 6.5 nJ

For a circular device with a radius of 5 μm, the device area is:

A = $\pi$ r² = pi × (5 μm)² = 78.5 μm²

Therefore, the area normalized write energy is:

E_write/A = 6.5 nJ / 78.5 μm²

= 0.0828 nJ/μm²

= 82.8 pJ/μm²

Assuming the write current scales with device area, corresponding to an approximately constant switching current density during lateral scaling, the projected write energy would decrease substantially for nanoscale devices. Based on the area normalized value of ~82.8 pJ/μm² extracted from **Supplementary Fig. S16**. If the 50 nm node is defined by a 50 nm diameter device, the projected area is ~0.00196 μm², corresponding to a projected write energy of ~0.16 pJ per pulse. These estimates highlight the strong benefit of area scaling for ultrathin graded AlScN devices and suggest that nanosecond scale, low voltage switching could be compatible with sub pJ class write energies in highly scaled memory cells.